\documentclass[11pt]{article}
\usepackage{graphicx,amsmath,amsfonts,amssymb,bm,hyperref,url,breakurl,epsfig,epsf,color,fullpage,MnSymbol,mathbbol, fmtcount,algorithmic,algorithm, semtrans
} 
\usepackage{subfigure}
\usepackage{xcolor}
\newtheorem{theorem}{Theorem}[section]

\usepackage{hyperref}
\definecolor{darkred}{RGB}{150,0,0}
\definecolor{darkgreen}{RGB}{0,150,0}
\definecolor{darkblue}{RGB}{0,0,200}
\hypersetup{colorlinks=true, linkcolor=darkred, citecolor=darkgreen, urlcolor=black}

\title{Accelerated Wirtinger Flow:  A fast algorithm for ptychography}

\author{Rui Xu, Mahdi Soltanolkotabi, Justin P.~Haldar, Walter Unglaub,\\
Joshua Zusman, Anthony F.~J.~Levi, Richard M.~Leahy\\
Ming Hsieh Department of Electrical Engineering\\
 University of Southern California, Los Angeles, CA 90089}
\date{January 2018; Revised June 2018}
\begin{document}
\maketitle
\begin{abstract}
This paper presents a new algorithm, Accelerated Wirtinger Flow (AWF), for ptychographic image reconstruction from phaseless diffraction pattern measurements. AWF is based on combining Nesterov's acceleration approach with Wirtinger gradient descent.  Theoretical results enable prespecification of  all AWF algorithm parameters, with no need for computationally-expensive line searches and no need for manual parameter tuning. AWF is evaluated in the context of simulated X-ray ptychography, where we demonstrate fast convergence and low per-iteration computational complexity. We also show examples where AWF reaches higher image quality with less computation than classical algorithms.  AWF is also shown to have robustness to noise and probe misalignment.
\end{abstract}



\section{Introduction}



The ability to image objects at nm scales is of fundamental importance in a variety of scientific and engineering disciplines. For instance, successful imaging of large protein complexes and biological specimens at very fine scales may enable live imaging of biochemical behavior at the molecular level providing new insights.  Similarly, modern multilayered integrated circuits increasingly contain features below 10nm in size. The ability to image such specimens non-destructively can  be used to improve quality control during the manufacturing process. 

Imaging at finer and finer resolutions necessitates shorter and shorter beam wavelengths. However, lens-like devices and other optical components are difficult to build at very short wavelengths. Phase-less coherent diffraction imaging techniques offer an alternative method for recovery of high resolution images without the need for involved measurment setups that include mirrors and lenses. Invention of new light sources and new experimental setups that enable recording and reconstruction of non-crystalline objects has caused a major revival in the use of phase-less imaging techniques \cite{miao1999, shapiro2005biological, chapman2006high, williams2006fresnel, pfeifer2006three, abbey2008keyhole, nishino2009three, nelson2010high, jiang2010quantitative, seibert2011single, clark2013ultrafast, xu2013single, gallagher2014macromolecular}. More recently, successful experiments using ptychography \cite{rodenburg2007hard, thibault2008high, dierolf2010ptychographic, shapiro2014chemical, deng2015simultaneous, holler2017high}, Fourier ptychography \cite{zheng2013wide, williams2014fourier, horstmeyer2015digital, chung2015counting, tian2015computational, tian20153d} and partially coherent Ptychography \cite{chang2018partially} have further contributed to this surge. There has also been tremendous progress in the development of phase retrieval methods with the introduction of new algorithmic approaches that include maximum-likelihood estimation \cite{thibault2012}, Ptychographic Iterative Engine (PIE) \cite{rodenburg2004phase} and extended Ptychographic Iterative Engine (ePIE) \cite{maiden2009improved}, Difference Map (DM) \cite{thibault2009probe, thibault2008high}, Error Reduction (ER) \cite{yang2011iterative}, Relaxed Averaged Alternating Reflections (RAAR) \cite{marchesini2016sharp}, semidefinite programming \cite{candes_strohmer, CESV12, candes2014phase, jaganathan2016stft, udellsketchy}, convex relaxation with an anchor vector \cite{goldstein2018phasemax, bahmani2016phase, dhifallah2018phase, ghods2018phaselin, salehi2018precise}, eigen-based angular synchronization \cite{Alexeev, iwen2016phase}, Wirtinger Flow (WF) \cite{WF},  proximal algorithms \cite{wen2012,hesse2015,soulez2016}, and majorize-minimize methods \cite{weller2015}. See also  \cite{chen2015solving, sun2016geometric, soltanolkotabi2017structured,  zhang2016provable, bendory2018non, zhang2016reshaped, cai2015optimal, wang2016solving, ma2018optimization, qu2017convolutional, DavisEtAl2017nonsmooth, TanVershynin2017Phase, jeong2017convergence, chen2018gradient, waldspurger2016phase, duchi2017solving, chandra2017phasepack} for many interesting works on first-order methods and/or theoretical analysis with random sensing ensembles.


Despite all the aforementioned progress, major challenges impede the use of phaseless imaging techniques for imaging large specimens at nm scales. One major challenge is computational in nature. For example, imaging a 1 cm $\times$ 1 cm integrated circuit specimen at 10 nm resolution results in on the order of $10^{12}$ pixels that need to be reconstructed, and the image sizes grow dramatically when extended to 3D tomographic imaging applications.  Given the computational complexity of processing such large data, it is important that phase retrieval algorithms converge quickly to good solutions.  

For practical phaseless imaging modalities such as X-ray ptychography \cite{hegerl1970dynamische, rodenburg2004phase, rodenburg2007transmission, rodenburg2007hard}, classical phase retrieval algorithms often exhibit slow convergence rates. Another challenge is that the image reconstruction task in phaseless imaging often involves highly non-convex optimization problems with many spurious local optima so that classical approaches can converge to suboptimal solutions or even may not converge at all. Recently, theoretical results have proven that gradient-descent methods, also known as WFs, will converge with high probability to globally optimal solutions in the case of randomized sensing ensembles \cite{WF}. Random/psuedo-random sensing ensembles can be realized in the visible light range via spatial light modulators or phase from defocus \cite{jingshan2014transport}. Unfortunately, these theoretical results do not apply to the majority of real-world imaging scenarios currently used at nm scales. It is possible to implement randomized models such as low-pass coded diffraction patterns \cite{candes2014phase}, for example by moving a large sand sheet in front of the sample \cite{personcomm}. In the x-ray regime, a similar approach uses a sheet of paper to produce random structure in the illumination beam \cite{Stockmar2013}.  When used in practical applications like ptychography, these WF approaches also suffer from slow convergence rates, albeit to a lesser extent.

In this paper, inspired by a seminal acceleration technique by Nesterov from the optimization literature \cite{nesterov1983method} we present a new accelerated algorithm for phase retrieval called Accelerated Wirtinger Flow (AWF).  While use of Nesterov's method has previous been proposed in the context of phase retrieval\cite{zhou2016geometrical}, our approach differs in the specification of a fixed step-size based on a Lipschitz-like constant. We derive novel theoretical results that prove the convergence of WF algorithms to stationary points for arbitrary phase-less measurements. These theoretical guidelines allows us to to eliminate computationally-expensive line-search algorithms, which means that WF, and by extension AWF, has low per-iteration computational complexity and has no algorithm parameters to tune. We note that recently a parallel line of research utilizes acceleration in PIE-based algorithms \cite{maiden2017}. However, this approach diverges from Nesterov's formulation in several ways and still requires parameter tuning. 

Using X-ray ptychography simulations, we observe that our algorithm exhibits much faster convergence when compared to traditional WF and other popular algorithms from the literature. Our results also show that AWF can achieve higher quality image reconstruction while being significantly more efficient in terms of computation cost, and also demonstrates resilience in scenarios with noise and device misalignment. 


\section{Phase retrieval}
\subsection{Generic Phase Retrieval}
The phase retrieval problem arises in a variety of optical and X-ray imaging scenarios where the detector measures the intensity but not the phase of the diffraction field.  The general version of the acquisition model can be written as
\begin{equation}
\mathbf{d} = \left|\mathbf{A} \mathbf{f}_{*} \right|^2 + \mathbf{n},\label{eq:phase}
\end{equation}
where $\mathbf{d} \in \mathbb{R}^M$ is the vector of measured intensity values at the detector, $\mathbf{f}_{*} \in \mathbb{C}^N$ represents samples of the discretized complex-valued object function that we want to reconstruct, the matrix $\mathbf{A} \in \mathbb{C}^{M \times N}$ describes the propagation model  for the optical system, and $\mathbf{n} \in \mathbb{R}^M$ represents noise perturbations.  When applied to a matrix $\mathbf{A}$ or vector $\mathbf{a}$, we use the notation $\left|\mathbf{A}\right|$ or $\left|\mathbf{a}\right|$ to denote applying the absolute-value function to each entry elementwise, e.g., $\left[\left|\mathbf{A}\right|\right]_{mn} = \left|\left[\mathbf{A}\right]_{mn}\right|$ for each $m=1,2,\ldots, M$ and each $n=1,2,\ldots,N$, where $\left[\mathbf{A}\right]_{mn}$ is the entry from the $m$th row and $n$th column of $\mathbf{A}$.

\subsection{Ptychographic Phase Retrieval}
\begin{figure}[htbp]
\centering
\includegraphics[scale=0.44]{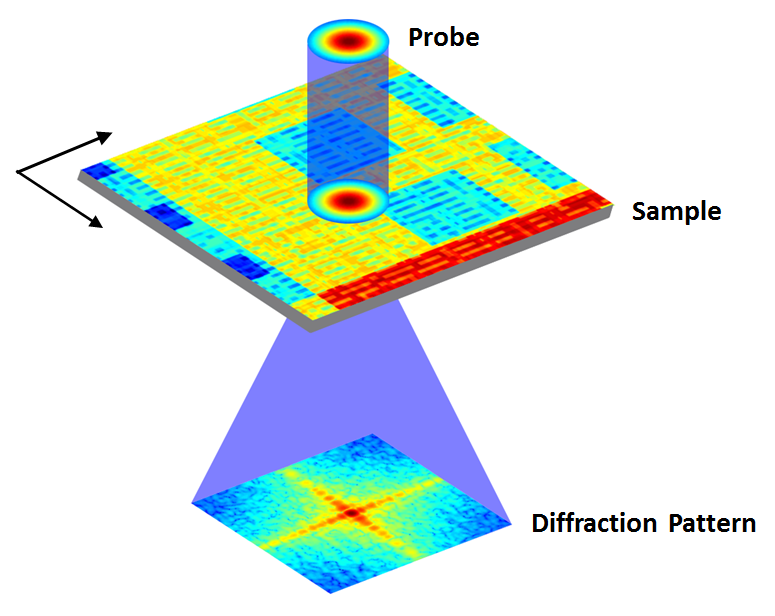}
\caption{A schematic illustration of ptychography.  A 2D sample is illuminated by a 2D Gaussian probe function, and a diffraction pattern is measured by a detector in the far field.}
\label{fig:schematic}
\end{figure}

Ptychography \cite{hegerl1970dynamische, rodenburg2004phase, rodenburg2007transmission, rodenburg2007hard} is a coherent diffractive imaging method that leads to a special case of the generic phase retrieval problem.  In this experiment, a sample is illuminated with several different illumination functions (or ``probes") and corresponding diffraction patterns for each probe are measured by a detector in the far field.  If $f(\mathbf{r})$ represents the 2D object function  (corresponding to $\mathbf{f}$ in Eq.~\eqref{eq:phase}) as a function of the spatial position $\mathbf{r}=(x,y)$ and $p_k(\mathbf{r})$ represents the $k$th probe function, then the complex field at the detector plane resulting from the $k$th probe is given by
\begin{equation}
b_k(\mathbf{r}') = \mathcal{F}\left\{p_k(\mathbf{r}) f(\mathbf{r}) \right\},\label{eq:cont}
\end{equation}
where $\mathcal{F}\{f(\mathbf{r})\}$ denotes the Fourier transform of $f(\mathbf{r})$ with respect to $\mathbf{r}$.  In many cases, the different probe functions $p_k(\mathbf{r})$ are obtained as different spatial shifts of the same basic probe function $p(\mathbf{r})$, i.e., $p_k(\mathbf{r}) = p(\mathbf{r}-\mathbf{r}_k)$ for some set of position vectors $\mathbf{r}_k$, $k=1,2,\ldots,K$.   A schematic illustration of ptychograhy is shown in Fig. 
.~\ref{fig:schematic}.

If data is collected for $K$ different probe functions, then the $\mathbf{A}$ matrix from Eq.~\eqref{eq:phase} corresponding to a discretization of Eq.~\eqref{eq:cont} can be written as
\begin{align}
\mathbf{A}=\begin{bmatrix}\mathbf{F} &\mathbf{0}&\mathbf{0} &\ldots &\mathbf{0}\\\mathbf{0}&\mathbf{F}&\mathbf{0} &\ldots& \mathbf{0}\\\mathbf{0}&\mathbf{0}&\mathbf{F}&\ldots&\mathbf{0}\\\vdots & \vdots & \vdots &\ddots &\vdots\\\mathbf{0}&\mathbf{0}&\mathbf{0}&\ldots&\mathbf{F}\end{bmatrix}\begin{bmatrix}\mathbf{I}_{\text{supp}(\mathbf{p}_1)}\text{diag}\left(\mathbf{p}_1\right)\\ \mathbf{I}_{\text{supp}(\mathbf{p}_2)}\text{diag}\left(\mathbf{p}_2\right)\\\vdots\\ \mathbf{I}_{\text{supp}(\mathbf{p}_K)}\text{diag}\left(\mathbf{p}_K\right)\end{bmatrix}.\label{eq:ptychmat}
\end{align}
Here, $\mathbf{p}_k \in \mathbb{C}^N$  represents samples of the discretized complex-valued $k$th probe function, diag$(\mathbf{p}_k)\in\mathbb{C}^{N\times N}$ is a diagonal matrix with diagonal entries equal to the entries of $\mathbf{p}_k$. We assume $\mathbf{p}_k$ has contiguous support of size $\ell$ and $\mathbf{I}_{\text{supp}(\mathbf{p}_k)}\in\mathbb{R}^{\ell\times N}$ is a sub-matrix of the $N\times N$ identity matrix $\mathbf{I}$, indexed by the set of non-zero entries of $\mathbf{p}_k$ or $\text{supp}(\mathbf{p}_k)$. Finally, $\mathbf{F}$ is an $M/K\times \ell$ matrix modeling the 2D Fourier transform and detector sampling operation and it is assumed that each of the $K$ diffraction patterns is measured at $M/K$ spatial positions in the detector plane. Typically, in ptychography $\ell=M/K$ and $\mathbf{F}$ is an $M/K\times M/K$ DFT matrix.

\section{Reconstruction Algorithm: Accelerated Wirtinger Flow}

\subsection{Basic Formulation as an Optimization Problem}
The AWF algorithm is designed to estimate $\mathbf{f}$ from Eq.~\eqref{eq:phase} by solving an optimization problem of the form 
\begin{equation}
\hat{\mathbf{f}} = \arg\min_{\mathbf{f} \in \mathbb{C}^N} \mathcal{L}(\mathbf{f}).\label{eq:opt}
\end{equation}
In this work, we choose $\mathcal{L}: \mathbb{C}^N\rightarrow \mathbb{R}$  to be the cost function
\begin{equation}
\mathcal{L}(\mathbf{f}) = \bigg\| \mathbf{b} - \left| \mathbf{A}\mathbf{f} \right| \bigg\|_2^2,\label{eq:cost}
\end{equation}
where $\mathbf{b}$ is obtained by taking the elementwise square-root of $\mathbf{d}$, i.e., $[\mathbf{b}]_m = \left([\mathbf{d}]_m\right)^{1/2}$ for $m=1,2,\ldots,M$, and the $\ell_2$-norm of the vector $\mathbf{a}\in\mathbb{C}^N$ is defined by $\|\mathbf{a}\|_2 = \left(\sum_{n=1}^N \left|[\mathbf{a}]_n\right|^2\right)^{1/2}$.  This choice of cost function, based on the amplitude rather than intensity of the diffraction pattern, can be derived as an approximation of the maximum likelihood (ML) cost function for a Poisson noise model \cite{thibault2012}, and can therefore be justified from the perspective of statistical estimation theory \cite{kay1993}.   In addition, we and many other authors \cite{thibault2012,soltanolkotabi2017structured,yeh2015experimental,zhang2016reshaped,wang2016solving,yeh2015experimental} have  observed that this choice of cost function leads to faster convergence behavior and better noise robustness compared to other cost functions (e.g., the ML cost function that would be obtained by assuming $\mathbf{n}$ in Eq.~\eqref{eq:phase} follows a Gaussian distribution, or the unapproximated ML cost function for the Poisson distribution). Since this behavior is well established in the literature  we focus on the cost \eqref{eq:cost} in the remainder of the paper and do not provide comparisons using other costs.

\subsection{A Review of Wirtinger Flow and some New Results}

The AWF algorithm is a generalization of the WF method \cite{WF}.  As a result, this subsection reviews the WF method, and also presents some useful new theory that can be used to significantly reduce the computational cost associated with each iteration of WF.  Later in the paper we will also make use of this computational improvement for AWF.

The optimization problem in Eq.~\eqref{eq:opt} is nonlinear and generally does not have a simple closed-form solution.  As a result, it is common to rely on iterative minimization methods.  Gradient descent is one of the simplest and most natural iterative minimization algorithms.  In this approach, starting from some initial guess $\hat{\mathbf{f}}_0$, the estimate of $\mathbf{f}$ is updated at the $\tau$th iteration according to  
\begin{equation}
\hat{\mathbf{f}}_{\tau+1}=\hat{\mathbf{f}}_\tau-\mu_\tau\nabla \mathcal{L}(\hat{\mathbf{f}}_\tau),\label{eq:iter}
\end{equation}
where $\mu_\tau$ is the step-size for the $\tau$th iteration, and $\nabla\mathcal{L}(\hat{\mathbf{f}}_\tau)$ is the gradient  of Eq.~\eqref{eq:cost} with respect to $\mathbf{f}$, evaluated at the point $\hat{\mathbf{f}}_\tau$.  Strictly speaking, Eq.~\eqref{eq:cost} is not complex-differentiable.  However, it is still possible to define a generalized gradient based on the notion of Wirtinger derivatives.  Following~\cite{WF}, we shall refer to the iterations defined by Eq.~\eqref{eq:iter} as WF.

The generalized gradient for Eq.~\eqref{eq:cost}  takes the form \cite{WF}
\begin{align}
\label{gengrad}
\nabla \mathcal{L}(\mathbf{f}):=\mathbf{A}^H\left(\mathbf{A}\mathbf{f}-\mathbf{b}\odot \mathrm{sgn}\left(\mathbf{A}\mathbf{f}\right)\right).
\end{align}
In this expression, we have used $\mathbf{A}^H$ to denote the conjugate transpose of the matrix $\mathbf{A}$, and have used $\odot$ to denote elementwise multiplication of two vectors (i.e., for vectors $\mathbf{a}$ and $\mathbf{b}$ in $\mathbb{C}^N$, we have that $\mathbf{a}\odot\mathbf{b}$ is also a length-$N$ vector defined by $\left[\mathbf{a} \odot \mathbf{b}\right]_n =  [\mathbf{a}]_n [\mathbf{b}]_n$ for $n=1,2,\ldots,N$).  We have also introduced the complex signum function, which is defined for vectors $\mathbf{a} \in \mathbb{C}^N$ as the length-$N$ vector that obeys
\begin{equation}
[\mathrm{sgn}(\mathbf{a})]_n = \left\{
\begin{array}{ll} 
0, & \text{if } [\mathbf{a}]_n = 0 \\ 
\frac{[\mathbf{a}]_n}{\left|[\mathbf{a}]_n\right|}, & \text{otherwise},
\end{array} 
\right. 
\end{equation}
for $n=1,2,\ldots,N$.

Given this expression for the gradient, it remains to choose the step size $\mu_\tau$ to complete the specification of the WF algorithm.  One of the most popular approaches to selecting $\mu_\tau$ is to solve a simple one-dimensional ``line-search" optimization problem  \cite{press1992}
\begin{equation}
\mu_\tau = \arg \min_{\mu \in \mathbb{R}} \mathcal{L}\left(\hat{\mathbf{f}}_\tau-\mu\nabla \mathcal{L}(\hat{\mathbf{f}}_\tau)\right).\label{eq:line}
\end{equation}
This choice is useful in the sense that it ensures the maximal possible decrease of the cost function along the gradient direction.  On the other hand, solving this optimization problem can also be computationally expensive.  Specifically in the case of ptychography with $\mathbf{A}$ defined in Eq.~\eqref{eq:ptychmat}, each cost function evaluation requires the computation of $K$ two-dimensional Fourier transform operations in addition to the $2K$ two-dimensional Fourier transform operations that are needed to compute the gradient vector.  Consequently, solving Eq.~\eqref{eq:line} can be computationally expensive if a large number of candidate $\mu$ values need to be evaluated.

Here we make the novel observation that the Wirtinger algorithm will always converge if we set $\mu_\tau$ to be the constant value
\begin{equation}
\mu_\tau = \bar{\mu} := \frac{1}{\lambda_{\mathrm{max}}(\mathbf{A}^H\mathbf{A})}\label{eq:step}
\end{equation}
for all $\tau$, where $\lambda_{\mathrm{max}}(\mathbf{A}^H\mathbf{A})$ is the largest eigenvalue of the positive semidefinite matrix $\mathbf{A}^H\mathbf{A}$.  Importantly, this observation means that WF can be implemented without a  line search.  The choice of this stepsize is justified by Theorem \ref{generalA} in the Appendix. This theorem demonstrates that with our chosen step size the Wirtinger Flow iterates converge to a point where the generalized gradient is zero. This is a non-trivial statement as the loss function is non-smooth and there can be many stationary points where the generalized gradient does not vanish. The main use of this theorem in this paper is to justify our choice of step size in Eq.~\eqref{eq:step}. 

Computing $\bar{\mu}$ is particularly straightforward when $\mathbf{A}$ represents the ptychographic propagation model.  Specifically, assuming that the $\mathbf{A}$ matrix has the form of Eq.~\eqref{eq:ptychmat}, then  $\mathbf{A}^H\mathbf{A}$ can be written as
\begin{equation}
\mathbf{A}^H\mathbf{A} = \sum_{k=1}^K \mathrm{diag}(\mathbf{p}_k)^H\mathbf{I}_{\text{supp}(\mathbf{p}_K)}^H \mathbf{F}^H \mathbf{F} \mathbf{I}_{\text{supp}(\mathbf{p}_K)}\mathrm{diag}(\mathbf{p}_k).
\end{equation}
In practical ptychography contexts, the Fourier transform operator $\mathbf{F}$ is often chosen in such a way that $\mathbf{F}^H\mathbf{F} = \alpha \mathbf{I}$ for some constant $\alpha$, where $\mathbf{I}$ is the $N\times N$ identity matrix.  In such cases, the previous expression simplifies to
\begin{equation}
\mathbf{A}^H\mathbf{A} = \alpha \sum_{k=1}^K \mathrm{diag}(\mathbf{p}_k)^H  \mathbf{I}_{\text{supp}(\mathbf{p}_K)}^H\mathbf{I}_{\text{supp}(\mathbf{p}_K)}\mathrm{diag}(\mathbf{p}_k)= \alpha \sum_{k=1}^K \mathrm{diag}(\mathbf{p}_k)^H \mathrm{diag}(\mathbf{p}_k).\label{eq:ptychmu}
\end{equation}
The right-hand side of Eq.~\eqref{eq:ptychmu} is the sum of $K$  diagonal matrices, and is therefore also diagonal.  Since the eigenvalues of a diagonal matrix are equal to its diagonal entries, it is easy to see that $\lambda_{\mathrm{max}}(\mathbf{A}^H\mathbf{A})$ for this case is equal to the maximum entry of the easy-to-compute length-$N$ vector $ \alpha \sum_{k=1}^K |\mathbf{p}_k| \odot |\mathbf{p}_k|$.  

For other choices of loss function in \eqref{eq:opt} (in lieu of \eqref{eq:cost}), it may be possible to solve the exact line search problem \eqref{eq:line} in closed form. For instance, the cost function that minimizes the least-squares fit on intensity rather than amplitude values is a fourth order polynomial and hence \eqref{eq:line} can be solved by cubic rooting \cite{jiang2016wirtinger, wei2017conjugate}. However, as mentioned previously we and other authors have observed that the amplitude-based loss \eqref{eq:cost} leads to faster convergence behavior and better noise robustness compared to other loss functions (even when exact line search is used). We are not aware of any closed form solutions to \eqref{eq:line} for the cost function \eqref{eq:cost} that is the focus of this paper.

\subsection{From Wirtinger Flow to Accelerated Wirtinger Flow}
While the WF method described in the previous subsection often demonstrates faster convergence behavior than other classical phase retrieval schemes, the convergence rate can still be rather slow. To overcome this challenge,  we replace the WF iteration from Eq.~\eqref{eq:iter} with the following AWF update equation:
\begin{align}
\hat{\mathbf{f}}_{\tau+1}=\hat{\mathbf{f}}_\tau+\beta_\tau\left(\hat{\mathbf{f}}_\tau-\hat{\mathbf{f}}_{\tau-1}\right)-\bar{\mu}\nabla \mathcal{L}\left(\hat{\mathbf{f}}_\tau+\beta_\tau\left(\hat{\mathbf{f}}_\tau-\hat{\mathbf{f}}_{\tau-1}\right)\right),\label{eq:awf}
\end{align}
with $\bar{\mu}$ defined in Eq.~\eqref{eq:step}, $\nabla\mathcal{L}\left(\mathbf{f}\right)$ defined in Eq.~\eqref{gengrad}, and $\beta_\tau$ defined by
\begin{align}
\label{params}
\beta_\tau:=\frac{\tau+1}{\tau+3}.
\end{align}
It should be noted that, aside from the choice of initialization $\hat{\mathbf{f}}_0$ and the choice of stopping criterion, the AWF algorithm is fully specified in Eq.~\eqref{eq:awf}. It has no tuning parameters and requires no line searches.  The computational cost of evaluating each iterative update using Eq.~\eqref{eq:awf} is also low, and approximately the same as the computational cost of computing the gradient vector.  

This AWF algorithm is inspired by the seminal work of Nesterov \cite{nesterov1983method}, who showed that if $\mathcal{L}(\mathbf{f})$ is smooth and convex, then the iterations defined by Eq.~\eqref{eq:awf} converge at an optimal rate.  However, the cost function in \eqref{eq:cost} is neither convex nor differentiable, and we have not proven that AWF will always converge to a minimizer of Eq.~\eqref{eq:cost}. Therefore, it may not be immediately clear that our proposed AWF scheme is useful, although many other authors have empirically observed that applying Nesterov-inspired algorithms to nonconvex cost functions can lead to outstanding results \cite{ghadimi2016accelerated}. Our numerical evaluations below suggest that in practice, AWF often converges much more rapidly than both WF and classical phase retrieval algorithms. 

We note that Zhou, Zhang, and Liang \cite{zhou2016geometrical} also study accelerated phase retrieval using Nesterov's approach. However this paper differs from ours in terms of the problem setting and conclusions. In particular, this paper does not specify an appropriate step-size for ptychography, uses a different acceleration parameter, develops local convergence guarantees, and is focused on Gaussian designs in lieu of ptychography. We emphasize that utilizing acceleration for nonconvex optimization is of course well established. The novelty of our approach is three-fold: (1) specification of a step-size in terms of the Lipschitz-like constant for this problem, (2) demonstrating that such an approach works empirically for the practically relevant ptychography model without any need for tuning the step size, and (3) demonstrating that this approach can in some cases even escape shallow local minima.

\section{Numerical Experiments}
\label{secnum}

In this section, we investigate the performance of AWF in the context of ptychographic phaseless imaging.  This work was motivated by x-ray ptychographic imaging of integrated circuits at resolution on the order of 10nm using a synchronton x-ray source\cite{holler2017high}, and we have designed our simulations in accordance with this application.

\subsection{Simulated Ptychography Experimental Setup}
\begin{figure}[htbp]
\centering
\subfigure[Probe Function, $p(\mathbf{r})$]{\includegraphics[width=0.45\textwidth]{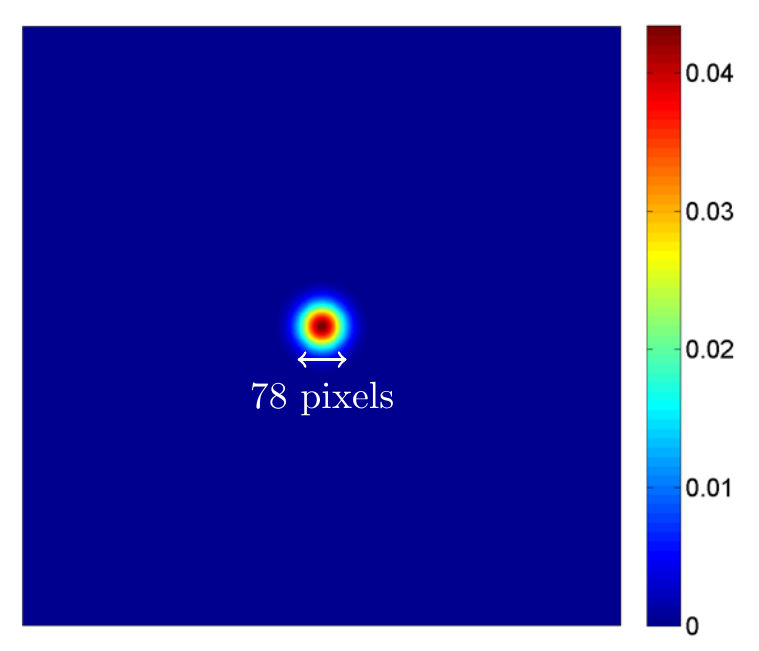}}
\subfigure[Illumination Pattern]{\includegraphics[width=0.45\textwidth]{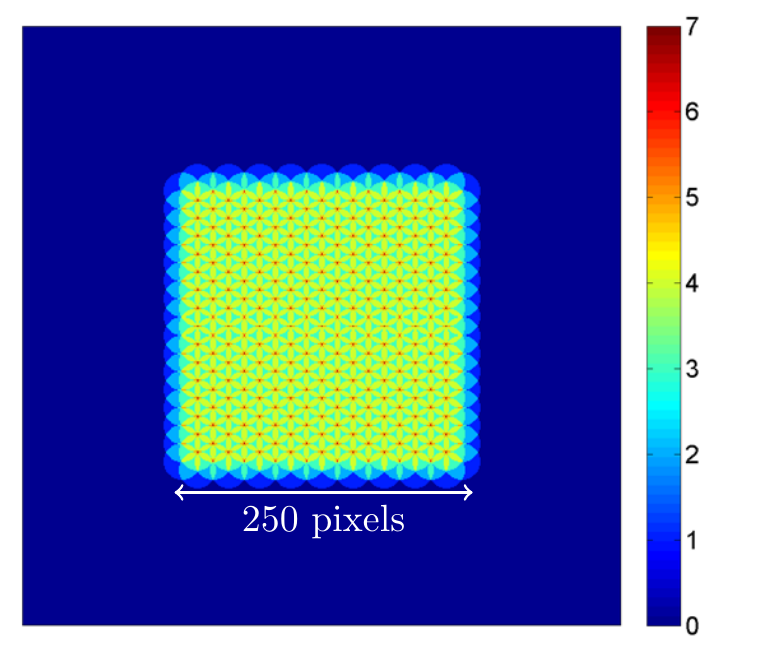}}
\caption{Images depicting (a) the probe function and (b) the illumination pattern used in the nominal simulated ptychography experiments.}\label{2}
\end{figure}

We consider complex test images (to be described later) with $500 \times 500$ pixels (i.e., $N=250,000$).  To mimic the true illumination in a physical experiment, we use multiple shifts of the basic probe function $p(\mathbf{r})$, and set $p(\mathbf{r})$ to  be a 2D Gaussian function with full width at half maximum (FWHM) of 30 pixels as shown in Fig. \ref{2}(a). The probe is set to zero outside of the central 78 pixel $\times$ 78 pixel region. 

In the ideal noiseless case, this probe is shifted to $K=313$ positions $\mathbf{r}_k$ based on a hexgonal scan pattern over a $250\times 250$ portion at the center of the sample. The distance between two adjacent scan spots is 15 pixels, 50\% of the FWHM. We have depicted the overall illumination pattern in Fig. \ref{2}(b) where the pixel values indicate how many times each pixel is illuminated. To prepare a stack of $313$ diffraction patterns, we start from the first scan location, extract a $160 \times 160$ frame and multiply it by the probe centered on this frame. We then compute a 2D Discrete Fourier Transform (DFT) of the result. The use of 160 pixels in the frame determines the sample interval in Fourier space. The magnitude of the transform is recorded and saved before we move to the next scan location. In such a setting $M/K = $25,600 corresponding to $M = 8,012,800$.

In addition to the ideal noiseless case, we also considered practical scenarios with either Poisson noise resulting from a limited number of photons or probe position misalignment introduced by imperfections in either the beam or sample stage positioning hardware.
We simulate psuedo-Poisson noise based on the assumption that  either $1 \times 10^9$ or $1 \times 10^7$ photons are delivered to the sample for each diffraction pattern. To model misalignment, we used a random displacement for each $\mathbf{r}_k$ probe position value when generating data (but used the nominal $\mathbf{r}_k$ values in reconstruction).  To do this we generated random independent displacement values along $x$ and $y$ by drawing uniformly from the set of possible displacements  $\{-1, 0, 1\}$ (in units of voxels) in each direction for each probe position.

\subsection{Description of Test Images}

Reconstructions were performed using two different test images. one that we numerically simulated based on the structure of an integrated circuit (IC), and the other a popular ptychography test image of gold beads.

\begin{figure}[htbp]
\centering 
\subfigure[Magnitude]{\includegraphics[width=0.4\textwidth]{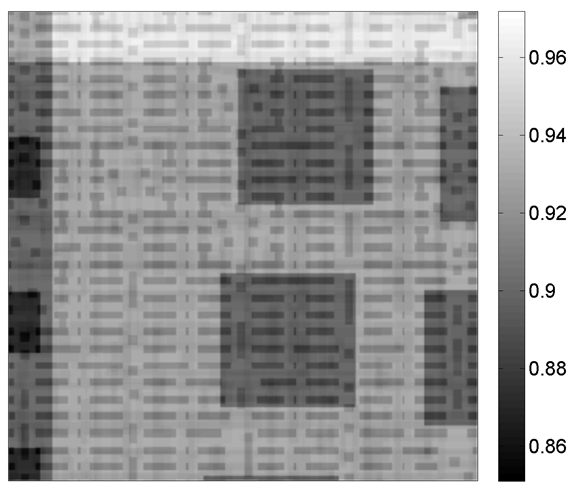}}
\subfigure[Phase (degrees)]{\includegraphics[width=0.4\textwidth]{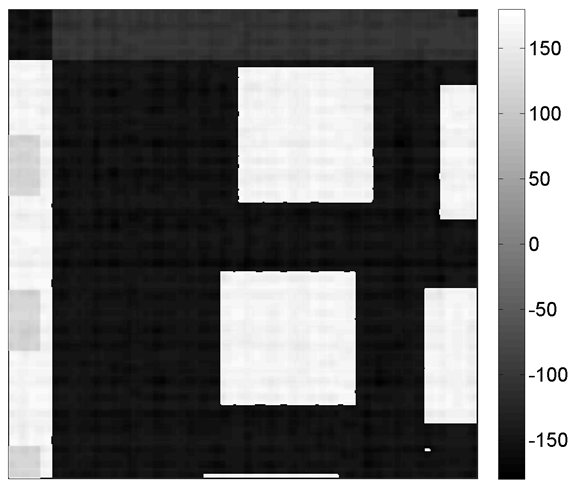}}
	\caption{Magnitude and phase of the integrated circuit test image.}
	\label{sampIC}
\end{figure}

The IC test image is generated from the projection of a synthetic IC structure measuring 2.5 $\mu$m $\times$ 2.5 $\mu$m $\times$ 4.4 $\mu$m with a voxel size of 5 nm, corresponding to 500 $\times$ 500 $\times$ 880 voxels. This IC structure contains silicon dioxide (SiO$_{2}$), silicon (Si), aluminum (Al), tungsten (W), and copper (Cu), and these materials are modeled as complex dielectrics discretized in space, described by complex 
 $n_{index} = 1-\delta + i\beta$ that depends on the illumination energy of the x-ray probe; $\delta$ is the refractive index decrement responsible for shifting the probe phase, while $\beta$ is the imaginary part that describes attenuation of the probe amplitude by the material. We use a simulated x-ray source with an energy of 6.2 keV ($\lambda_{0} = 0.2$ nm) to simulate operating conditions of recent experiments \cite{dierolf2010ptychographic,vila2011characterization,holler2014x}. The test image, Fig.~\ref{sampIC},  is computed as the complex projection through the chip along the $z$-axis:
\begin{align}
T(x,y)=e^{\int_{0}^{\Delta z}ik_{0}n_{index}(x,y,z;k_{0})dz},
\end{align}
where $k_{0} = 2\pi/\lambda_{0}$ is the x-ray wavenumber and the integral is calculated over the chip thickness $\Delta z = 4.4\ \mu$m, resulting in a 500 $\times$ 500 complex-valued image. At an x-ray energy of 6.2 keV, Si has  values $\delta = 1.285 \times 10^{-5}$ and $\beta = 4.777 \times 10^{-7}$; Al has  values $\delta = 1.436 \times 10^{-5}$ and $\beta = 4.305 \times 10^{-7}$; W has  values $\delta = 7.971 \times 10^{-5}$ and $\beta = 9.840 \times 10^{-6}$; Cu has values $\delta = 4.264 \times 10^{-5}$ and $\beta = 1.479 \times 10^{-6}$; and SiO$_{2}$ has  values $\delta = 1.206 \times 10^{-5}$ and $\beta = 2.574 \times 10^{-7}$.

\begin{figure}[htbp]
\centering 
\subfigure[Magnitude]{\includegraphics[width=0.4\textwidth]{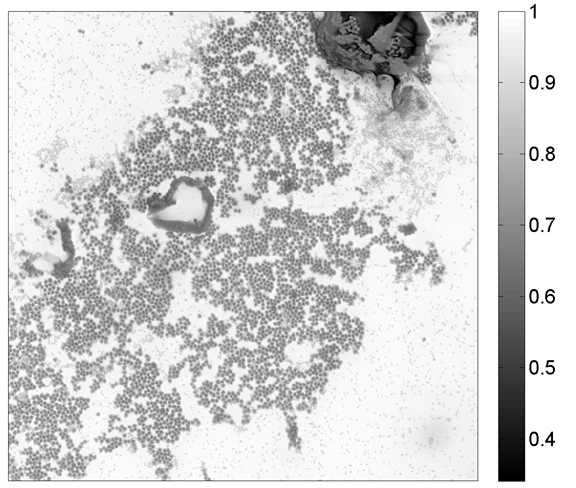}}
\subfigure[Phase (degrees)]{\includegraphics[width=0.4\textwidth]{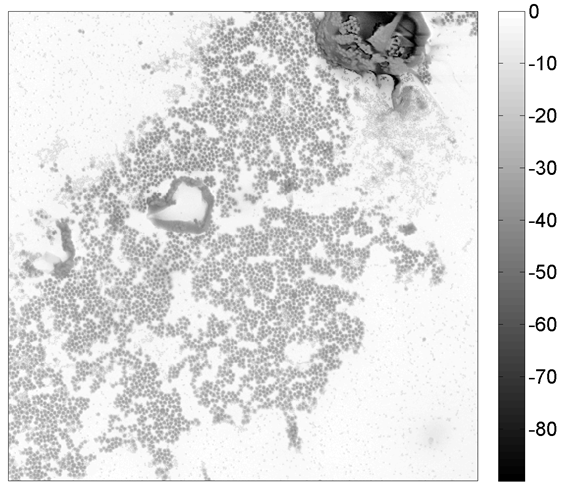}}
\caption{Magnitude and phase of the gold bead test image.}
\label{sampGB}
\end{figure}

The second test image is based on the transmission function of a collection of gold beads deposited on a membrane \cite{yang2011iterative}. The magnitude and phase of this test image are depicted in Fig.~\ref{sampGB}. 

\subsection{Comparison of Ptychography Algorithms.}
Images were reconstructed using our novel AWF algorithm and several previously-proposed ptychography algorithms for comparison.  Specifically, we implemented WF \cite{WF}, the Polak-Ribiere form of the nonlinear  Conjugate Gradient Method (CGM) \cite{thibault2012,yang2011iterative,press1992}, DM \cite{thibault2009probe, thibault2008high}, ePIE  \cite{maiden2009improved}, RAAR \cite{marchesini2016sharp}, and ER \cite{yang2011iterative}.  DM, ePIE, RAAR and ER are alternating projections algorithms that do not use the cost function from Eq.~\eqref{eq:cost}, while WF, AWF, and CGM are specifically designed to minimize Eq.~\eqref{eq:cost}.  To ensure a fair comparison, all seven algorithms are initialized in exactly the same way by setting $[\hat{\mathbf{f}}_0]_n=1$ for all $n=1,\ldots,N$.

Our WF and CGM implementations incorporate a line-search procedure to pick the step-size $\mu_\tau$ as in Eq.~\eqref{eq:line}.  Our implementation of this line search uses MATLAB's default line search algorithm (golden section search with parabolic interpolation \cite{forsythe1977computer}).

Our primary measure of the quality of the reconstructed images uses the normalized root mean-squared error between the ground truth and reconstructed images after compensating for phase ambiguities and masking the central portion of the image:
\begin{align}
\text{Relative reconstruction error} = \underset{\phi\in[0,2\pi]}{\min}\text{ }\frac{\left\|\mathbf{M}\left(e^{i\phi}\hat{\mathbf{f}}_\tau-\mathbf{f}_*\right)\right\|_2}{\left\|\mathbf{M}\mathbf{f}_*\right\|_2}.\label{eq:error}
\end{align}
The diagonal matrix $\mathbf{M}\in \mathbb{R}^{N/4 \times N}$ has the effect of extracting the central $250 \times 250$ pixel region of the reconstructed image and discarding the other regions, and is used because the other image regions received very little illumination and are not expected to be reconstructed accurately.    Our error metric has a phase correction factor $\phi$ because we only have magnitude measurements, which means that $\mathcal{L}(\hat{\mathbf{f}}) = \mathcal{L}(e^{i\phi}\hat{\mathbf{f}})$ for any choice of $\phi$. Therefore, the solution to Eq.~\eqref{eq:opt} is almost never unique, and it is only possible to recover the original signal/image up to a global phase factor. 

To investigate the convergence behavior of different algorithms, we compute and plot the error metric from Eq.~\eqref{eq:error} for each iteration up to $\tau=1000$.  This measure ignores variations in per-iteration computation time among algorithms. For example, multiple iterations of a line search can result in significantly higher costs per iteration. True computational cost is highly dependent on the specifics of each implementation and the computing platform. To avoid a potentially unfair comparison, we therefore also plot the error metric as a function of the number of FFTs used. This serves as a surrogate for computation cost since the measure is robust to implementational details and FFTs are the dominant computational burden in ptychographic image formation.  We also investigate the convergence of the cost function itself. In this case we only plot the behavior for three of the algorithms since only WF, AWF and CGM explicitly optimize the cost function in Eqs.  \ref{eq:opt}and \ref{eq:cost}. As with the image error metric, we plot the change in cost function as a function of number of iterations and number of FFTs.

\begin{figure}[htbp]
\centering
\subfigure[]{\includegraphics[scale=0.825]{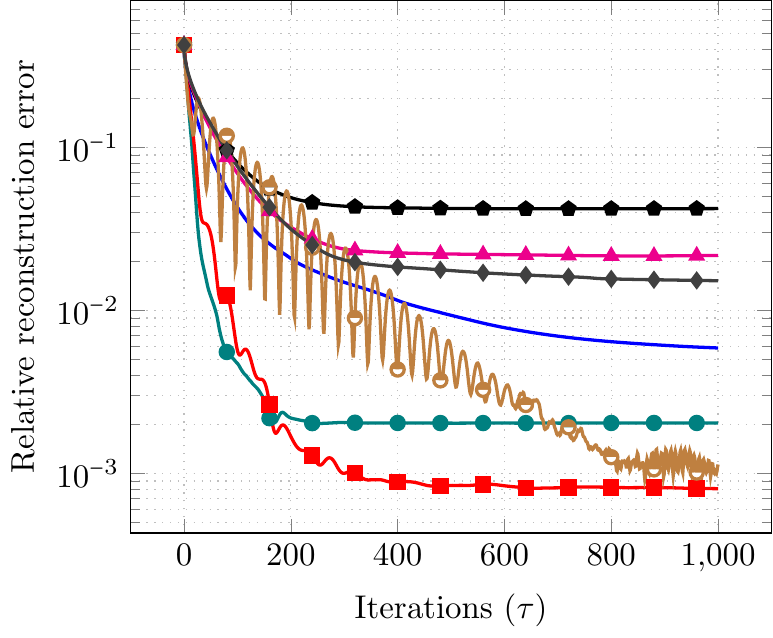}}
\hfill 
\subfigure[]{\includegraphics[scale=0.825]{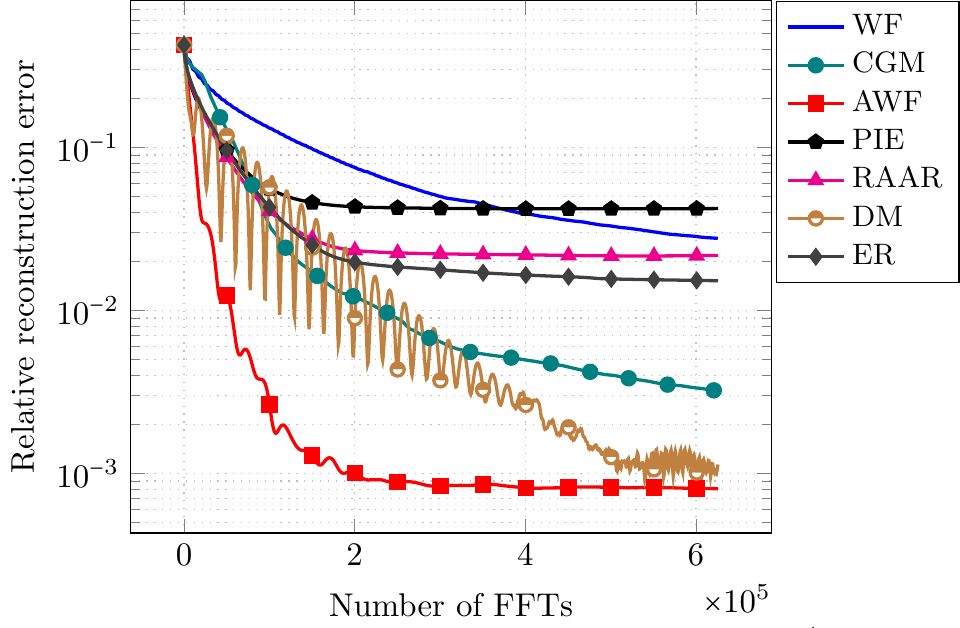}}
\subfigure[]{\includegraphics[scale=0.825]{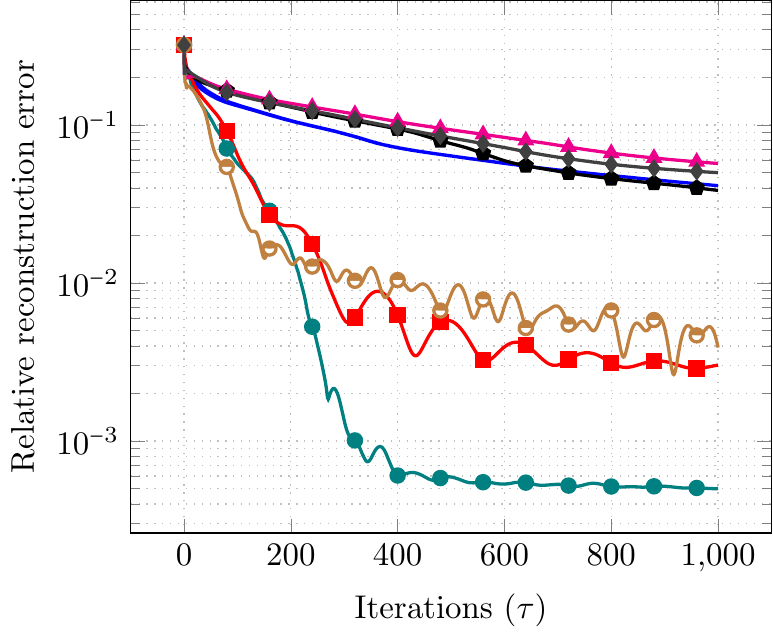}}
\hfill 
\subfigure[]{\includegraphics[scale=0.825]{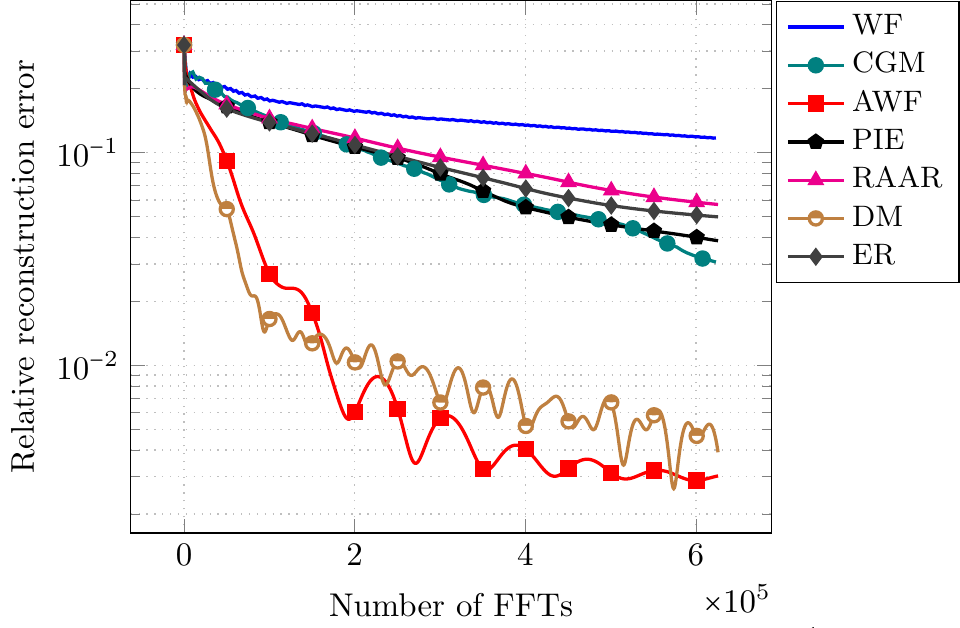}}
\caption{Convergence of the relative reconstruction error (Eq. \ref{eq:error}) for different algorithms applied to noiseless data from: (a,b) IC and (c,d) Gold Bead test images, plotted as a function of (a,c) iteration number and (b,d) number of FFT computations. }
\label{ICnoiseless}
\end{figure}

In most pytchographic applications the probe may not be known exactly (although in many cases an accurate estimate of the probe is available or can be estimated directly from data). To investigate the  AWF algorithm when the probe is not known exactly we compare its performance to the e-PIE algorithm. To run AWF when the probe is not known we first initialize the probe (as discussed next) and then in each iteration perform the following two updates: (1) a signal updates using the AWF iterates assuming a known probe and (2) a probe update using e-PIE's probe update strategy. We investigate two initialization scenarios: (1) a \emph{deterministic} initialization where we use the inverse Fourier transform of the average of the diffraction patterns and (2) a \emph{random perturbation} where we add a small amount of Gaussian noise to the original probe and use that as our initialization. In both cases we normalized the initial estimate of the probe so it has total energy one (the sum of squares of the absolute value of entries is set to one). The former method is a standard practical initialization. While the latter initialization is not achievable in practice, the goal of studying this initialization is to demonstrate the performance of the algorithms in the presence of small and random uncertainty in the probe. We evaluate the performance of the algorithms using the following error metric
\begin{align}
\text{Relative reconstruction error} = \underset{c\in\mathbb{C}}{\min}\text{ }\frac{\left\|\mathbf{M}\left(c\hat{\mathbf{f}}_\tau-\mathbf{f}_*\right)\right\|_2}{\left\|\mathbf{M}\mathbf{f}_*\right\|_2}.\label{eq:error2}
\end{align}
This is analogous to \eqref{eq:error2}. The only difference is that we use a complex scalar $c\in\mathbb{C}$ instead of the global phase factor $e^{i\phi}$ to account for the inherent ambiguity in the scale of the image when the probe is not known since the measurements are formed from the product of these two unknown quantities.
\subsection{Numerical Results}

Fig. \ref{ICnoiseless} shows convergence behavior of the relative reconstruction error (RRE) (Eq. \ref{eq:error}) in the noiseless case for the IC and Gold Bead test images.  The plots on the left show the change in RRE vs number of iterations. Those on the right are individually rescaled in the horizontal axis to reflect number of FFT computations. Note that the curves on the right show behavior for different maximum numbers of iterations for some of the methods, depending on the number of FFTs required per iteration. There are two interesting aspects to these plots: (a) the convergence behavior varies markedly across methods and (b) there are substantial differences across methods in the RRE even after 1,000 iterations. This latter observation reflects, in part, differences in convergence rate, but also the fact that not all are converging towards the same solution. Since ER, DM, RAAR and PIE are not explicitly optimizing a cost function it is not suprising that they would not converge to the same solution. However, there are also significant differences in RRE for WF, AWF and CGM, all of which optimize the same cost function. In that case, the differences may be largely attributable to different convergence rates, although it is also possible that, since the cost function is not convex, they may be converging towards different solutions. 

The AWF method shows substantially faster convergence in RRE vs. iteration than simple WF, even though the former uses a fixed step size while the latter uses a line search. For this reason, the differences also appear even larger when RRE is plotted against number of FFTs. These differences reflect the well known accelaration associated with Nesterov-like iterations compared to steepest descent. Conjugate gradient also exhibits fast convergence relative to most other methods, in fact even faster than AWF in the case of the gold bead test image, Fig. \ref{ICnoiseless}(c). However, when we account for the additional cost of the line search by plotting convergence vs. number of FFTs, CGM becomes relatively slow leaving AWF as the most rapidly converging algorithm. Note also that DM exhibits rapid convergence in these noiseless data. However, as we see below, DM becomes unstable in the presense of noisy or inconsistent data.     

\begin{figure}[htbp]
\centering
\subfigure[]{\includegraphics[scale=0.825]{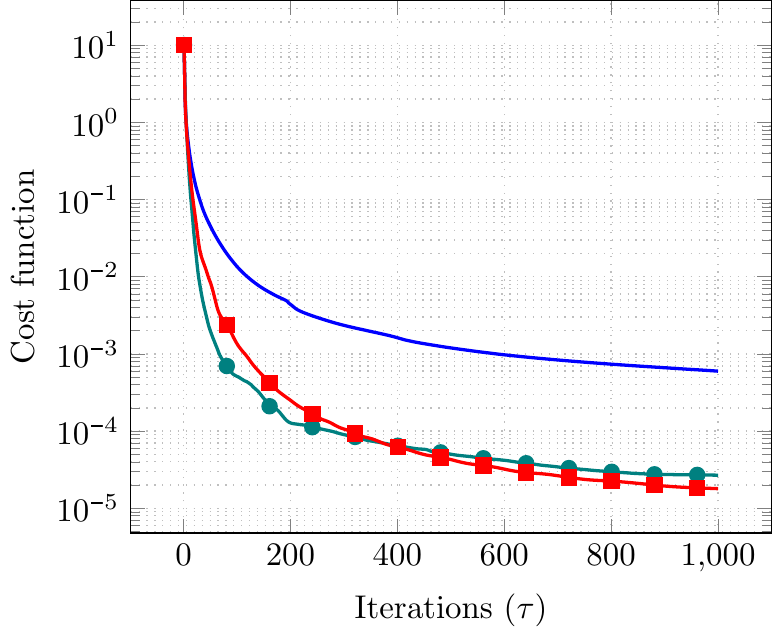}}
\hfill 
\subfigure[]{\includegraphics[scale=0.825]{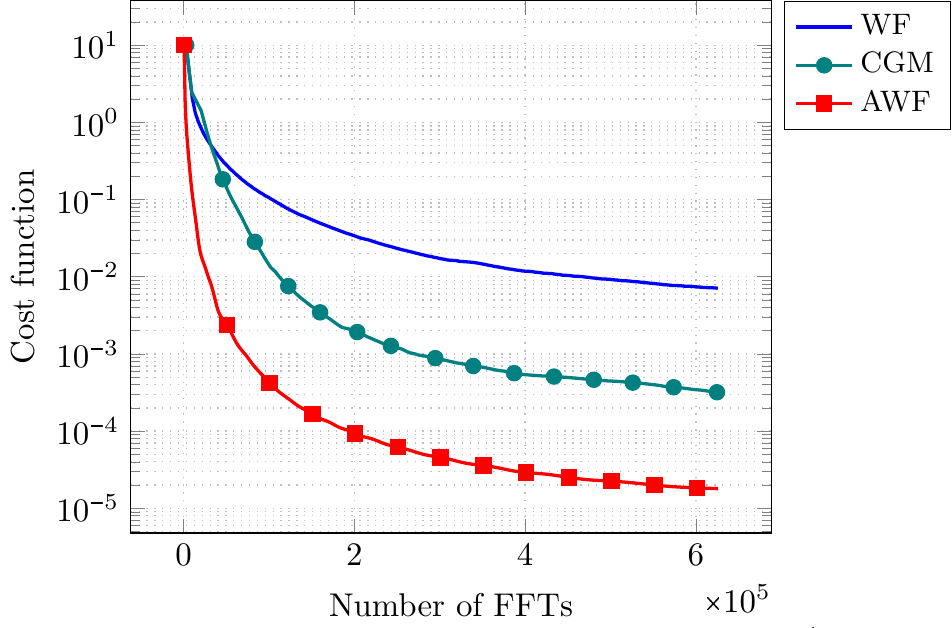}}
\subfigure[]{\includegraphics[scale=0.825]{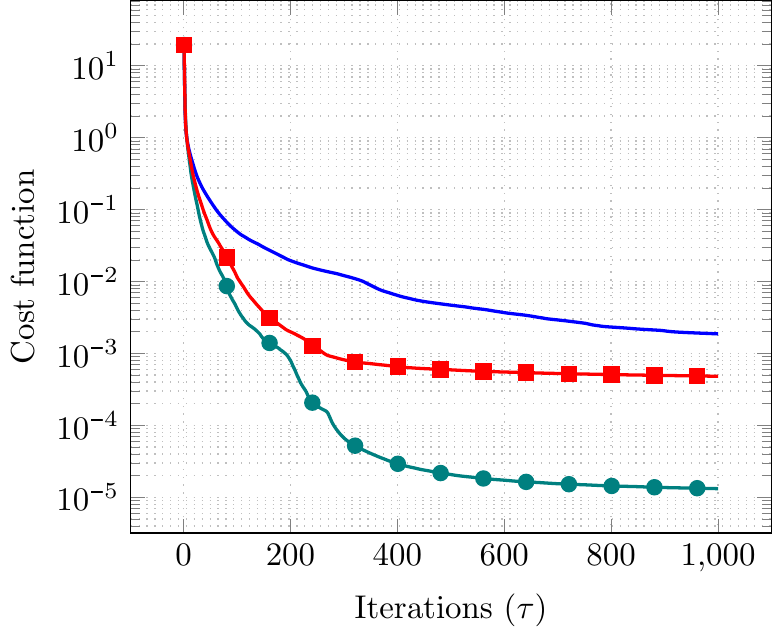}}
\hfill 
\subfigure[]{\includegraphics[scale=0.825]{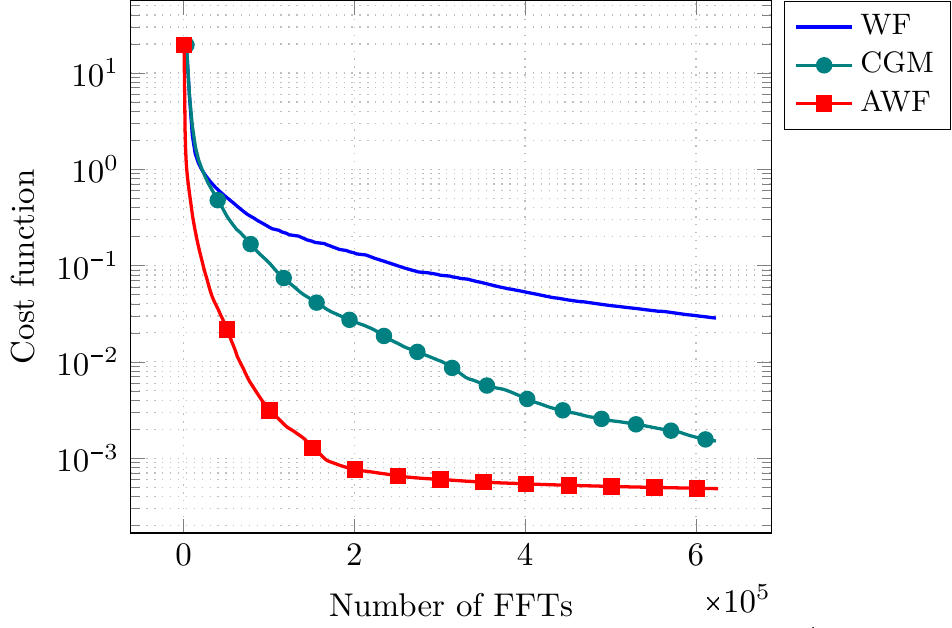}}
\caption{Convergence of cost function in Eqs. \ref{eq:opt}and \ref{eq:cost} for CGM, WF and AWF algorithms for noiseless data from (a,b) IC and (c,d) Gold Beads test images as a function of (a,c) iterations and (b,d) FFT computations. }
\label{ICcostfunction}
\end{figure}

Fig. \ref{ICcostfunction} shows the behavior of WF, AWF and CGM in terms of the cost function rather than the image error. Similar observations can be made with respect to their relative behavior. Although convergence rate for CGM can be fast as a function of iteration number, Fig. \ref{ICcostfunction}(a,c), once the additional costs of the line search is taken into account, AWF is substantially faster Fig. \ref{ICcostfunction}(b,d).

\begin{figure}[htbp]
\centering
\subfigure[]{\includegraphics[scale=0.825]{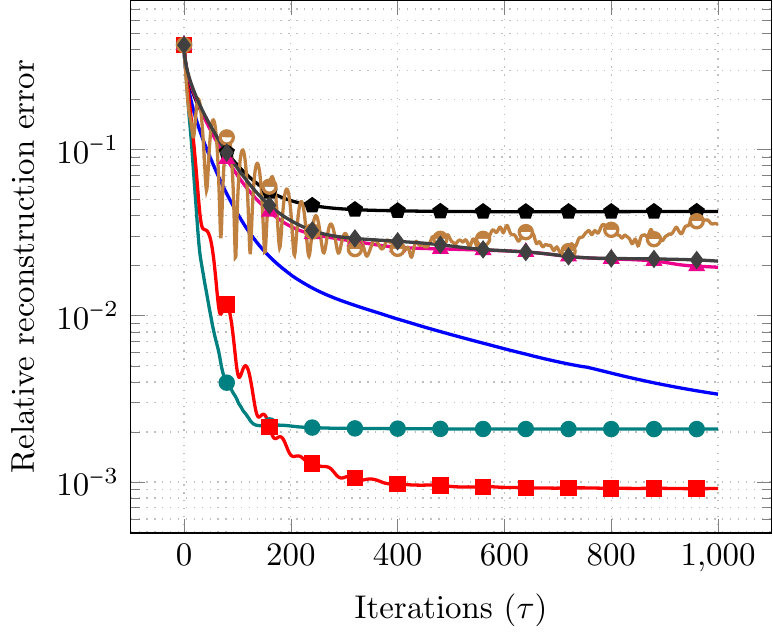}}
\hfill 
\subfigure[]{\includegraphics[scale=0.825]{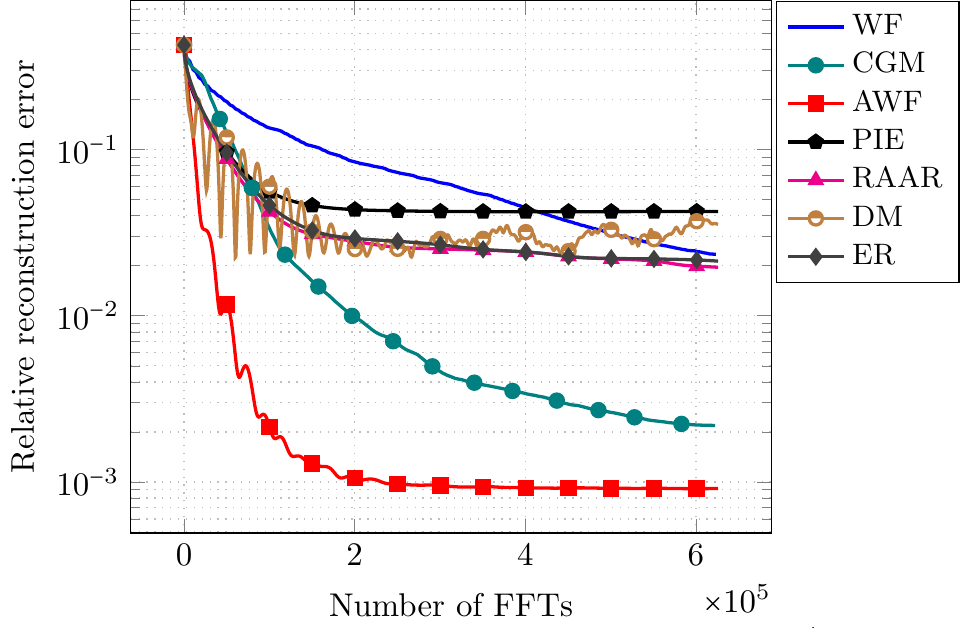}}
\subfigure[]{\includegraphics[scale=0.825]{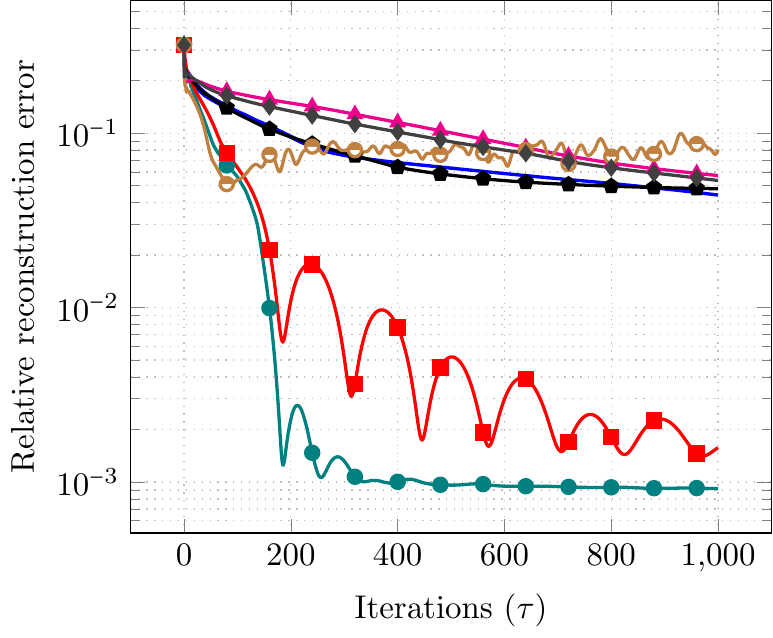}}
\hfill 
\subfigure[]{\includegraphics[scale=0.825]{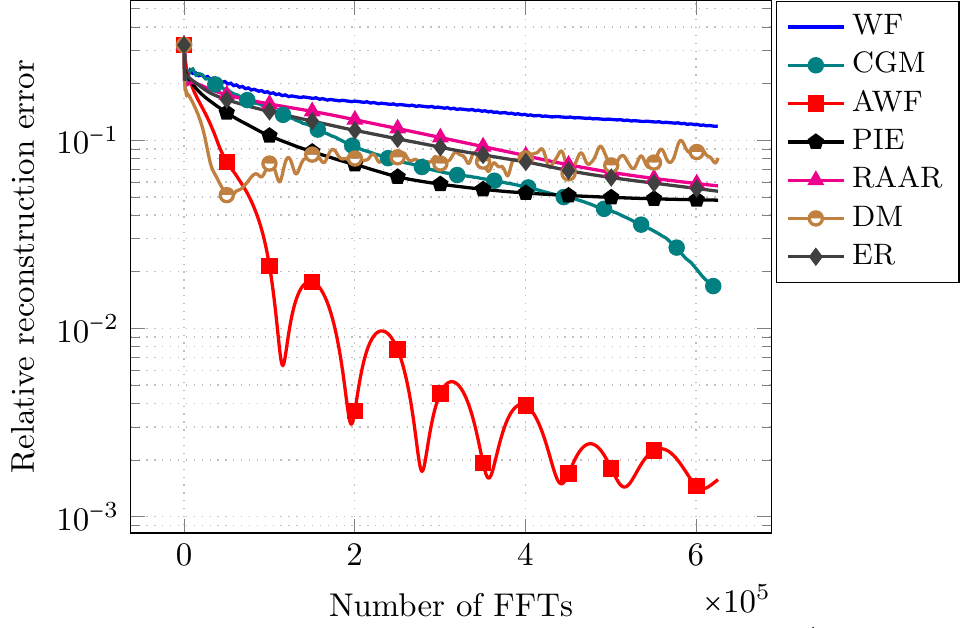}}
\caption{Convergence of the relative reconstruction error (Eq. \ref{eq:error}) for different algorithms applied to data with Poisson noise from: (a,b) IC and (c,d) Gold Bead test images, plotted as a function of (a,c) iteration number and (b,d) number of FFT computations.}
\label{noisyIC}
\end{figure}

\begin{figure}
\centering
\subfigure[]{\includegraphics[scale=0.6]{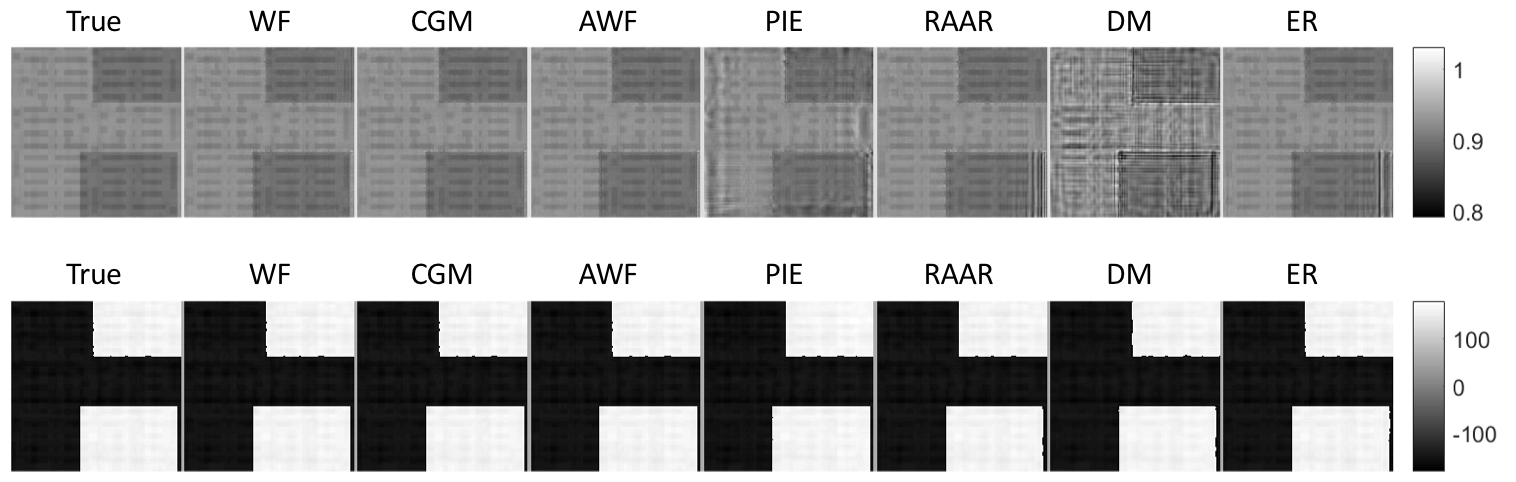}}
\hfill 
\subfigure[]{\includegraphics[scale=0.8]{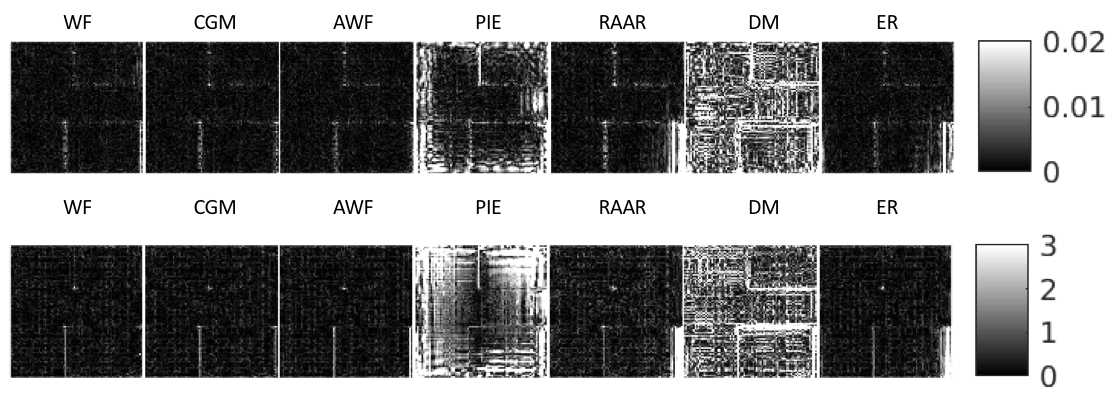}}
\caption{Magnitude (upper) and phase (lower) of (a) the ground truth and reconstructed images from different algorithms (b) the absolulute value of the magnitude and phase difference images (phase adjusted reconstructed image minus ground truth)  for IC sample from Poisson data, $10^7$ photons per probe position.}
\label{imagesIC1E7}
\end{figure}

\begin{figure}
\centering
\subfigure[]{\includegraphics[scale=0.6]{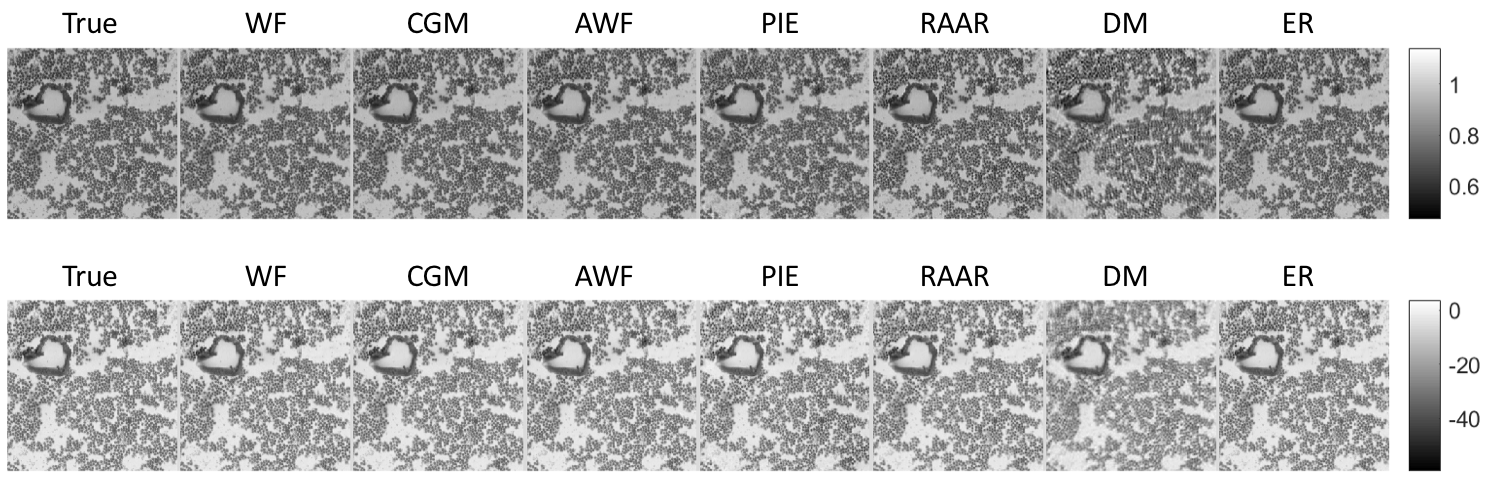}}
\hfill 
\subfigure[]{\includegraphics[scale=0.8]{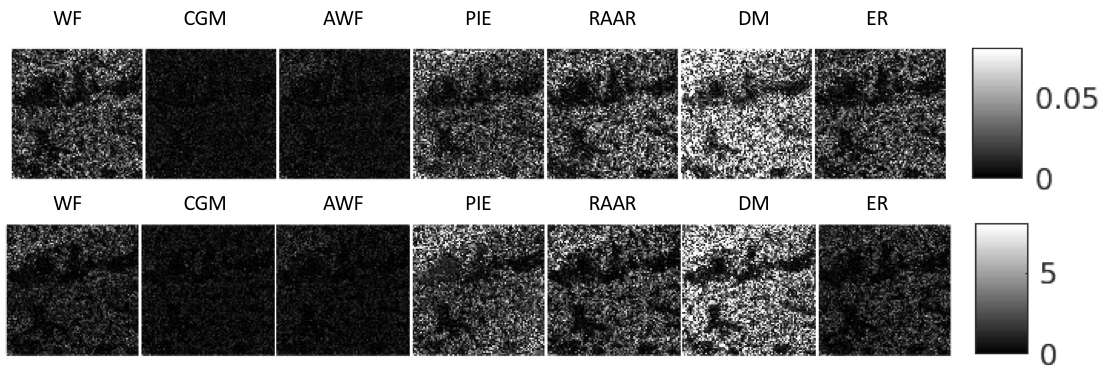}}
\caption{Magnitude (upper) and phase (lower) of (a) the ground truth and reconstructed images from different algorithms (b) the absolulute value of the magnitude and phase difference images (phase adjusted reconstructed image minus ground truth) for the golden bead sample from Poisson data, $10^7$ photons per probe position.}
\label{imagesGB1E7}
\end{figure}

\begin{figure}[htbp]
\centering
\subfigure[]{\includegraphics[scale=0.825]{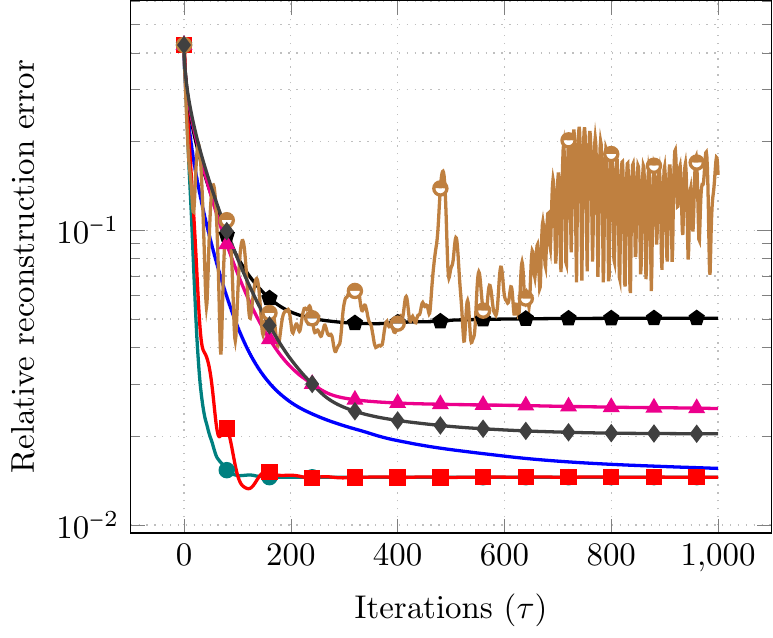}}
\hfill 
\subfigure[]{\includegraphics[scale=0.825]{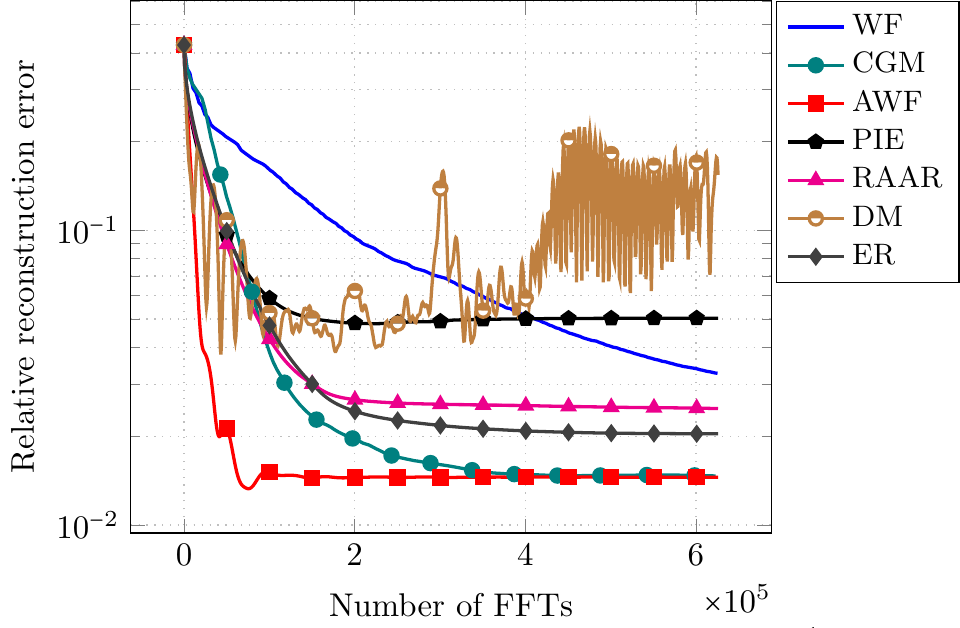}}
\subfigure[]{\includegraphics[scale=0.825]{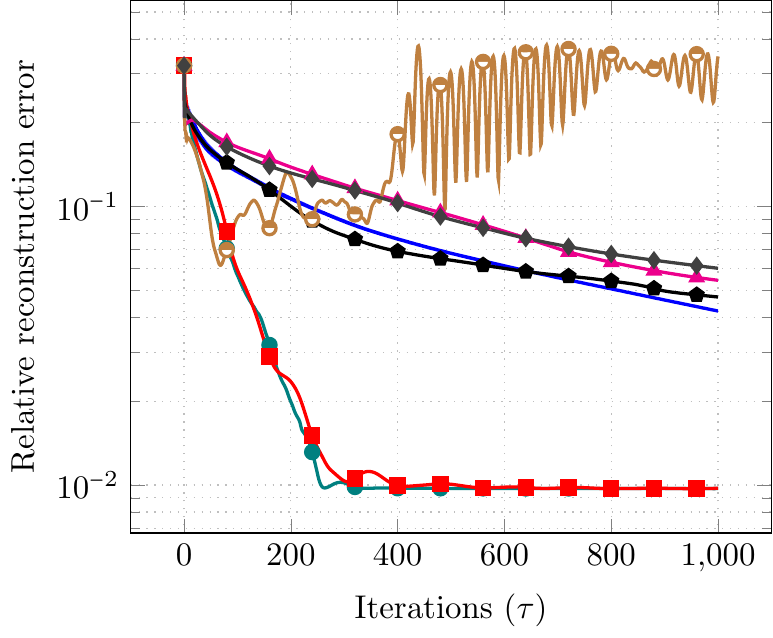}}
\hfill 
\subfigure[]{\includegraphics[scale=0.825]{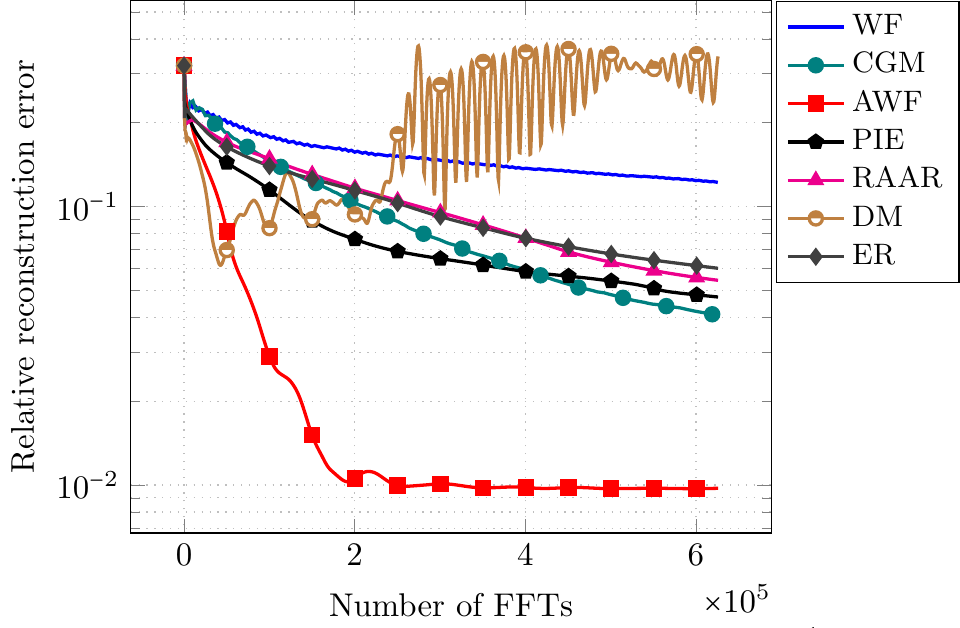}}
\caption{Convergence of the relative reconstruction error (Eq. \ref{eq:error}) for different algorithms applied to misaligned data from: (a,b) IC and (c,d) Gold Bead test images, plotted as a function of (a,c) iteration number and (b,d) number of FFT computations. }
\label{misIC}
\end{figure}

Fig. \ref{noisyIC} shows convergence behavior in the noisy case ($10^9$ photons per probe position) for the IC and Gold Bead test images while Fig. \ref{misIC} shows behavior when the data are misaligned. Figs. \ref{imagesIC1E7} and \ref{imagesGB1E7} show the 256 x 256 pixel central portion of the magnitude and phase of the ground truth images as well as those of the reconstructed images along with the absolute value of the magnitude and phase difference images (phase adjusted reconstructed images minus ground truth image) for the case of $10^7$ photons per probe position. We observe that most algorithms demonstrate similar characteristics to those that were seen in the noiseless case.   The major exception is the DM algorithm which exhibits quite unstable behavior in both cases. Consistently in these results, AWF and CGM exhibit the fastest convergence relative to iteration number. With the benefit of fewer FFTs required per iteration for AWF relative to CGM, the former shows substantially faster convergence when plotted against number of FFTs. Ther reconstructed images are visually similar for most cases, although errors are significant for DM. Overall, errors are larger in the magnitude than the phase component. Errors are consistently smaller for the optimization based methods (WF, CGM and AWF) than for PIE, RAAR, DM and ER. Interestingly, in some instances the errors in Fig. \ref{noisyIC} are smaller than for the noiseless case, Fig.  \ref{ICnoiseless}. We note that because the cost functions is non-convex these algorithms can become trapped in local optima and it is possible that a small amount of additional noise in the data may result in convergence to a different minimum that represents a better solution. 

\begin{figure}[htbp]
\centering
\subfigure[]{\includegraphics[scale=0.825]{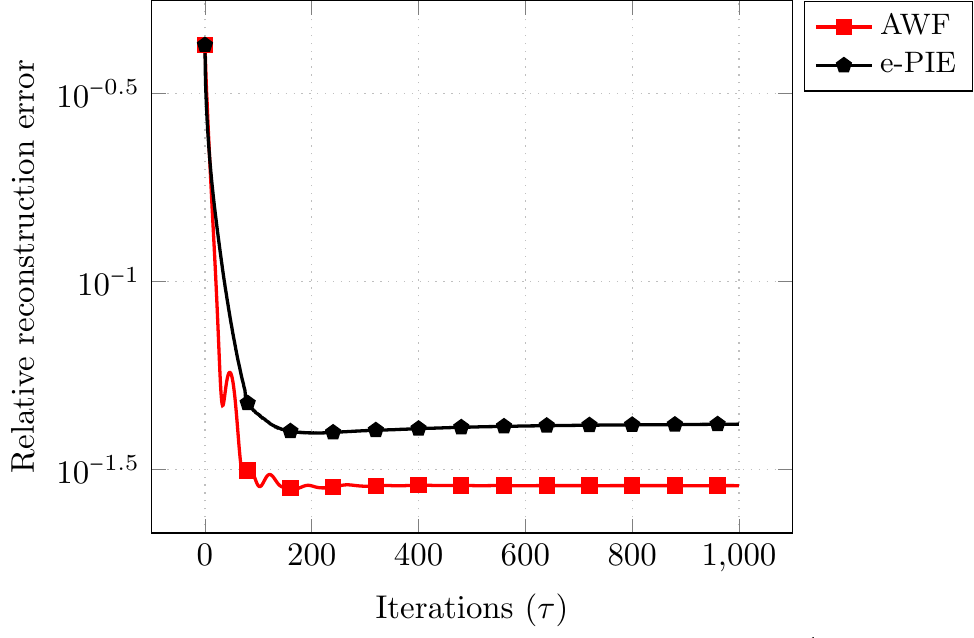}}
\subfigure[]{\includegraphics[scale=0.825]{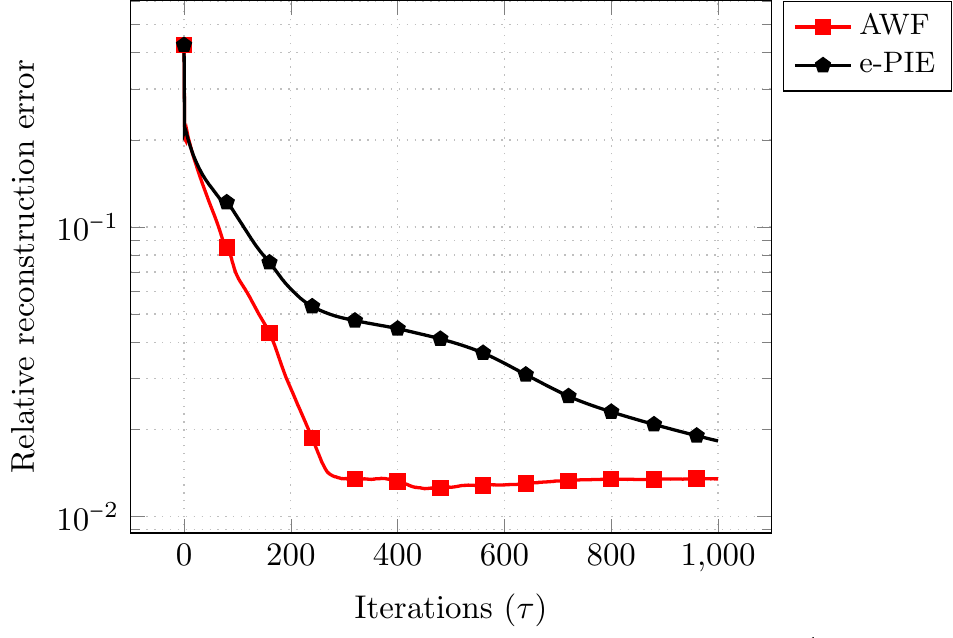}}
\caption{Convergence of the relative reconstruction error (Eq. \ref{eq:error2}) for e-PIE and AWF (using e-PIE's probe update applied to (a) IC and (b) Gold Bead test images, ploted as a function of iteration number. In this plot, the initial probe is set to a random Gaussian perturbation of the probe for both algorithms.}
\label{probeGauss}
\end{figure}

\begin{figure}[htbp]
\centering
\subfigure[]{\includegraphics[scale=0.825]{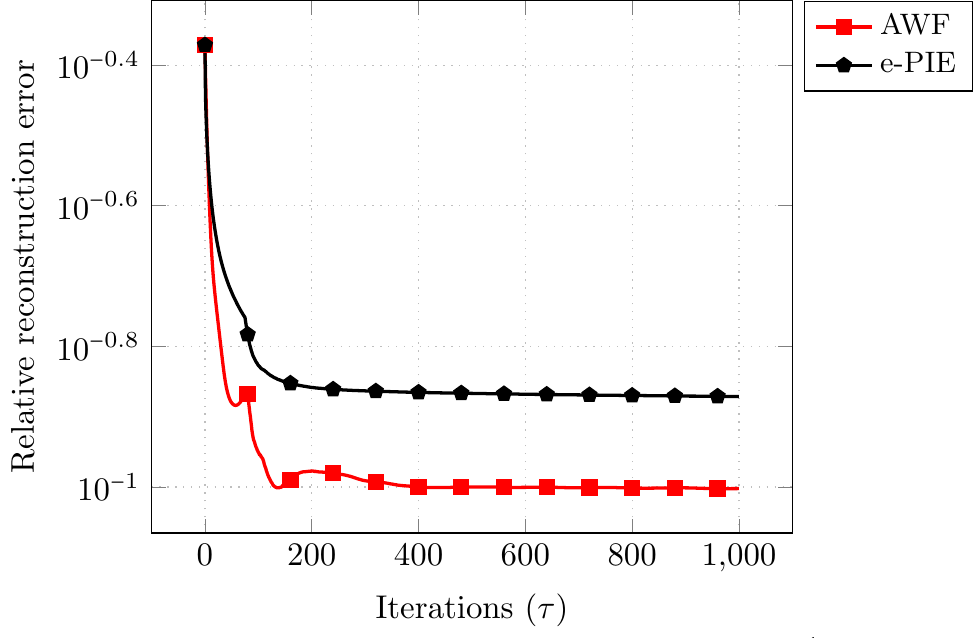}}
\subfigure[]{\includegraphics[scale=0.825]{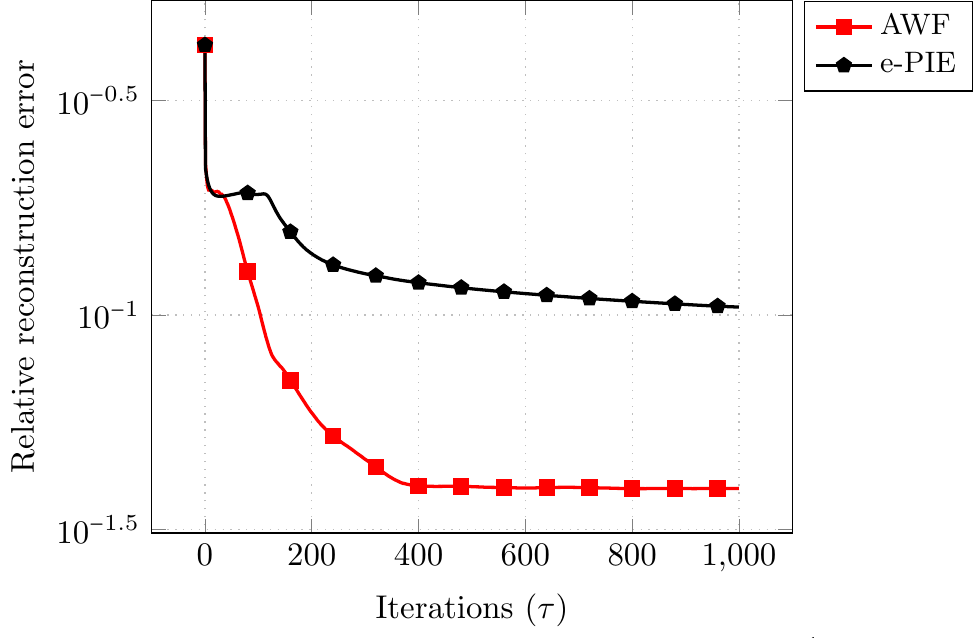}}
\caption{Convergence of the relative reconstruction error (Eq. \ref{eq:error2}) for e-PIE and AWF (using e-PIE's probe update applied to (a) IC and (b) Gold Bead test images, ploted as a function of iteration number. In this plot, the initial probe is set to the inverse Fourier transform of the average of the diffraction patterns for both algorithms.}
\label{probedeter}
\end{figure}

Fig. \ref{probeGauss} shows convergence behavior of AWF and e-PIE with probe updates at each iteration when they are both initialized via the random perturbation model for the noisy case ($10^9$ photons per probe position) for the IC and Gold Bead test images. Fig. \ref{probedeter} shows this convergence behavior when both algorithms are initialized via the deterministic initialization described in the previous section. We observe that in both cases, while errors are larger than in the case when the probe is known exactly, AWF converges faster to a smaller error compared to e-PIE.

\section{Discusion and Conclusions}
We have described the novel AWF algorithm for phase retrieval and demonstrated it on simulated ptychography data. AWF combines the Wirtinger flow approach to phase retrieval with Nesterov's momentum method for accelerated gradient descent. The primary advantage of AWF over alternative algorithms is that it explicitly optimizes a cost function without the need for a line search. We show that use of an easily computed Lipschitz-like constant ensures convergence of WF with a fixed step-size and also results in effective convergence in practical cases for AWF. Furthermore, computational results show favorable performance of AWF in terms of convergence rate, not only relative to WF and CGM (both of which optimize the same cost function) but also in comparison to other widely used ptychography methods. 

Interestingly, although the visible diffferences in the final images were relatively small, we do see significant differences in terms of residual errors, even after 1000 iterations of each algorithm, with either CGM or AWF consistently achieving the smallest error. Both of these fast algorithms are optimizing the same cost function so the similarity in performance when compared as a function of iteration is perhaps not surprising. When also accounting for the higher complexity in CGM as a result of the line-search, AWF exhibits the best combination of residual error and computation cost. WF also optimizes the same cost function but convergence is far slower. The other methods tested converge slower than AWF and use a range of alternation schemes which do not explicitly optimize a cost function and can result in differences in the final images with residual errors larger than AWF.    

Most of the simulation studies presented here assume that the probe is known exactly. However, we also show examples where we  update the probe estimate at each iteration using an e-PIE like procedure. These results shows that AWF retains its faster convergence behavior relative to e-PIE. These comparisons are preliminary in nature and are meant as a baseline comparison of AWF with a well known approach. We believe that eventually one may be able to develop an algorithm where the probe and signal are updated simultaneously using accelerated gradient updates. It may also be possible to develop rigorous theory for convergence to stationary points with a carefully developed step-size in this case. Such investigations are an interesting direction for future research


\section*{Acknowledgements}
This work supported in part by the Air Force Research Laboratory (AFRL)  under contract FA8650-17-C-9112. M. Soltanolkotabi is supported by the Air Force Office of Scientific Research Young Investigator Program (AFOSR-YIP) under
award number FA9550-18-1-0078 and a google faculty research award.


\bibliography{bibfiles}
\bibliographystyle{plain}

\appendix
\section*{APPENDIX}
\section{Theory for convergence to stationary points}
\label{sectheory}
He we present a theorem that provides theoretical justification for our choice of the step sizes \eqref{eq:step} and \eqref{params} for Wirtinger Flow and Accelerated Wirtinger Flow.

\begin{theorem}\label{generalA} Let $\mathbf{f}\in\mathbb{C}^N$ denote the desired signal and assume we have $M$ arbitrary noisy measurements of the form $\mathbf{b}=\left(\left|\mathbf{A}\mathbf{f}_{*}\right|^2+\mathbf{n}\right)^{1/2}\in\mathbb{R}^M$, where the square root is applied element-wise. Consider the cost function from Eq.~\eqref{eq:cost}.
We run the Wirtinger Flow updates from Eq.~\eqref{eq:iter} using the generalized gradient from Eq.~\eqref{gengrad} and the step size from Eq.~\eqref{eq:step}.  Also let $\hat{\mathbf{f}}\in\underset{\mathbf{f}\in\mathbb{C}^n}{\arg\min}$ $\mathcal{L}(\mathbf{f})$ be a global optima. Then, the following identities hold
\begin{align}
\label{statement}
\underset{\tau\rightarrow\infty}{\lim} \left\|\nabla \mathcal{L}(\hat{\mathbf{f}}_\tau)\right\|_2\rightarrow 0\quad\text{and}\quad \min_{\tau\in\{0,1,2,\ldots,T\}} \left\|\nabla \mathcal{L}(\hat{\mathbf{f}}_\tau)\right\|_2^2\le \frac{\left(\mathcal{L}(\hat{\mathbf{f}}_{0})-\mathcal{L}(\hat{\mathbf{f}})\right)}{\bar{\mu} (T+1)}.
\end{align}
\end{theorem}
\section{Proofs}
\label{secproofs}


We begin by studying the convergence of Wirtinger Flow iterates on a smoothed version of Eq.~\eqref{eq:cost}  defined by
\begin{align}
\mathcal{L}_{\epsilon}(\mathbf{f}):=&\sum_{m=1}^M \left(\left(\left|\mathbf{a}_m^H\mathbf{f}\right|^2+\epsilon\right)^{1/2}-[\mathbf{b}]_m\right)^2,
\end{align} 
where $\mathbf{a}_m^H \in \mathbb{C}^{1\times N}$ is the $m$th row of $\mathbf{A}$ and $\epsilon \in \mathbb{R}$ is a nonnegative scalar.  This cost function equals the cost function from Eq.~\eqref{eq:cost} when $\epsilon=0$, and is smooth when $\epsilon >0$.

To calculate Wirtinger derivatives we have to first rewrite $\mathcal{L}_{\epsilon}(\mathbf{f})$ as a holomorphic function of $\mathbf{f}$ and its conjugate $\bar{\mathbf{f}}$. For the loss above this takes the form
\begin{align*}
\mathcal{L}_{\epsilon}(\mathbf{f},\bar{\mathbf{f}}):=\mathcal{L}_{\epsilon}(\mathbf{f})=\sum_{m=1}^M \left(\left(\mathbf{f}^T\left(\mathbf{a}_m\mathbf{a}_m^H\right)^T\bar{\mathbf{f}}+\epsilon\right)^{\frac{1}{2}}-[\mathbf{b}]_m\right)^2
\end{align*}
Then the Wirtinger gradient can be calculated via the transpose of the partial derivative of $\mathcal{L}_{\epsilon}(\mathbf{f},\bar{\mathbf{f}})$ with respect to $\mathbf{f}$ with $\bar{\mathbf{f}}$ fixed (see \cite[Section 6]{WF} for background and further detail on Wirtinger gradient calculations). As a result for the loss considered in this paper the Wirtinger gradient is equal to
\begin{align*}
\nabla \mathcal{L}_{\epsilon}=\left(\frac{\partial }{\partial \mathbf{f}}\mathcal{L}_{\epsilon}(\mathbf{f},\bar{\mathbf{f}})\right)^T=\sum_{m=1}^M \frac{\left(\left(\mathbf{f}^T\left(\mathbf{a}_m\mathbf{a}_m^H\right)^T\bar{\mathbf{f}}+\epsilon\right)^{\frac{1}{2}}-[\mathbf{b}]_m\right)}{\left(\mathbf{f}^T\left(\mathbf{a}_m\mathbf{a}_m^H\right)^T\bar{\mathbf{f}}+\epsilon\right)^{\frac{1}{2}}}\left(\mathbf{a}_m\mathbf{a}_m^H\right)^T\bar{\mathbf{f}}.
\end{align*}

Also
\begin{align*}
\mathcal{H}_{\mathbf{f}\mathbf{f}}=&\frac{\partial }{\partial \mathbf{f}}\left(\frac{\partial \mathcal{L}_{\epsilon}}{\partial \mathbf{f}}\right)^H\\
=&\frac{\partial }{\partial \mathbf{f}}\left(\sum_{m=1}^M \frac{\left(\left(\mathbf{f}^T\left(\mathbf{a}_m\mathbf{a}_m^H\right)^T\bar{\mathbf{f}}+\epsilon\right)^{\frac{1}{2}}-[\mathbf{b}]_m\right)}{\left(\mathbf{f}^T\left(\mathbf{a}_m\mathbf{a}_m^H\right)^T\bar{\mathbf{f}}+\epsilon\right)^{\frac{1}{2}}}\left(\mathbf{a}_m\mathbf{a}_m^H\right)\mathbf{f}\right)\\
=&\sum_{m=1}^M \frac{\left(\sqrt{\left|\mathbf{a}_m^H\mathbf{f}\right|^2+\epsilon}-[\mathbf{b}]_m\right)}{\sqrt{\left|\mathbf{a}_m^H\mathbf{f}\right|^2+\epsilon}}\left(\mathbf{a}_m\mathbf{a}_m^H\right)+\frac{1}{2}\sum_{m=1}^M [\mathbf{b}]_m\frac{\left|\mathbf{a}_m^H\mathbf{f}\right|^2}{\left(\left|\mathbf{a}_m^H\mathbf{f}\right|^2+\epsilon\right)^{\frac{3}{2}}}\left(\mathbf{a}_m\mathbf{a}_m^H\right)\\
=&\sum_{m=1}^M\left(1-\frac{1}{2} \frac{[\mathbf{b}]_m}{\sqrt{\left|\mathbf{a}_m^H\mathbf{f}\right|^2+\epsilon}}-\frac{\epsilon}{2}\frac{[\mathbf{b}]_m}{\left(\left|\mathbf{a}_m^H\mathbf{f}\right|^2+\epsilon\right)^{\frac{3}{2}}}\right)\left(\mathbf{a}_m\mathbf{a}_m^H\right).
\end{align*}
\begin{align*}
\mathcal{H}_{\bar{\mathbf{f}}\mathbf{f}}=&\frac{\partial }{\partial \bar{\mathbf{f}}}\left(\frac{\partial \mathcal{L}_{\epsilon}}{\partial \mathbf{f}}\right)^H\\
=&\frac{\partial }{\partial \bar{\mathbf{f}}}\left(\sum_{m=1}^m \frac{\left(\sqrt{\bar{\mathbf{f}}^T\left(\mathbf{a}_m\mathbf{a}_m^H\right)\mathbf{f}+\epsilon}-[\mathbf{b}]_m\right)}{\sqrt{\bar{\mathbf{f}}^T\left(\mathbf{a}_m\mathbf{a}_m^H\right)\mathbf{f}+\epsilon}}\left(\mathbf{a}_m\mathbf{a}_m^H\right)\mathbf{f}\right)\\
=&\sum_{m=1}^M \frac{1}{2}\frac{[\mathbf{b}]_m}{\left(\left|\mathbf{a}_m^H\mathbf{f}\right|^2+\epsilon\right)^{\frac{3}{2}}}\left(\mathbf{a}_m^H\mathbf{f}\right)^2\mathbf{a}_m\mathbf{a}_m^T
\end{align*}
The Hessian takes the form
\begin{align*}
\nabla^2\mathcal{L}_{\epsilon}(\mathbf{f})=&\begin{bmatrix} \mathcal{H}_{\mathbf{f}\mathbf{f}}& \mathcal{H}_{\bar{\mathbf{f}}\mathbf{f}}\\\mathcal{H}_{\mathbf{f}\bar{\mathbf{f}}}&\mathcal{H}_{\bar{\mathbf{f}}\bar{\mathbf{f}}}\end{bmatrix}.
\end{align*}
We thus have
\begin{align*}
\begin{bmatrix}\mathbf{u}\\\bar{\mathbf{u}}\end{bmatrix}^H\nabla^2\mathcal{L}_{\epsilon}(\mathbf{f})\begin{bmatrix}\mathbf{u}\\\bar{\mathbf{u}}\end{bmatrix}=&\mathbf{u}^H\mathcal{H}_{\mathbf{f}\mathbf{f}}\mathbf{u}+\mathbf{u}^H\mathcal{H}_{\bar{\mathbf{f}}\mathbf{f}}\bar{\mathbf{u}}+\mathbf{u}^T\mathcal{H}_{\mathbf{f}\bar{\mathbf{f}}}\mathbf{u}+\mathbf{u}^T\mathcal{H}_{\bar{\mathbf{f}}\bar{\mathbf{f}}}\mathbf{u}\\
=&2\sum_{m=1}^M\left(1-\frac{1}{2} \frac{[\mathbf{b}]_m}{\sqrt{\left|\mathbf{a}_m^H\mathbf{f}\right|^2+\epsilon}}-\frac{\epsilon}{2}\frac{[\mathbf{b}]_m}{\left(\left|\mathbf{a}_m^H\mathbf{f}\right|^2+\epsilon\right)^{\frac{3}{2}}}\right)\left|\mathbf{a}_m^H\mathbf{u}\right|^2\\
&+\sum_{m=1}^M \frac{[\mathbf{b}]_m}{\left(\left|\mathbf{a}_m^H\mathbf{f}\right|^2+\epsilon\right)^{\frac{3}{2}}}\Re\left(\left(\mathbf{a}_m^H\mathbf{f}\right)^2(\mathbf{u}^H\mathbf{a}_m)^2\right)\\
=&2\sum_{m=1}^M\left(1-\frac{1}{2} \frac{[\mathbf{b}]_m\left|\mathbf{a}_m^H\mathbf{f}\right|^2}{\left(\left|\mathbf{a}_m^H\mathbf{f}\right|^2+\epsilon\right)^{\frac{3}{2}}}-\epsilon\frac{[\mathbf{b}]_m}{\left(\left|\mathbf{a}_m^H\mathbf{f}\right|^2+\epsilon\right)^{\frac{3}{2}}}\right)\left|\mathbf{a}_m^H\mathbf{u}\right|^2\\
&+\sum_{m=1}^M \frac{[\mathbf{b}]_m}{\left(\left|\mathbf{a}_m^H\mathbf{f}\right|^2+\epsilon\right)^{\frac{3}{2}}}\Re\left(\left(\mathbf{a}_m^H\mathbf{f}\right)^2(\mathbf{u}^H\mathbf{a}_m)^2\right)\\
=&2\sum_{m=1}^M\left(1-\epsilon\frac{[\mathbf{b}]_m}{\left(\left|\mathbf{a}_m^H\mathbf{f}\right|^2+\epsilon\right)^{\frac{3}{2}}}\right)\left|\mathbf{a}_m^H\mathbf{u}\right|^2\\
&+\sum_{m=1}^M \frac{[\mathbf{b}]_m}{\left(\left|\mathbf{a}_m^H\mathbf{f}\right|^2+\epsilon\right)^{\frac{3}{2}}}\left(\Re\left(\left(\mathbf{a}_m^H\mathbf{f}\right)^2(\mathbf{u}^H\mathbf{a}_m)^2\right)-\left|\mathbf{a}_m^H\mathbf{f}\right|^2\left|\mathbf{a}_m^H\mathbf{u}\right|^2\right)\\
\le&2\sum_{m=1}^M\left|\mathbf{a}_m^H\mathbf{u}\right|^2\\
\le&2\left\|\mathbf{A}\right\|^2\left\|\mathbf{u}\right\|_2^2\\
=&\left\|\mathbf{A}\right\|^2\left\|\begin{bmatrix}\mathbf{u}\\\bar{\mathbf{u}}\end{bmatrix}\right\|_2^2\\
=&\left\|\mathbf{A}^H\mathbf{A}\right\|\left\|\begin{bmatrix}\mathbf{u}\\\bar{\mathbf{u}}\end{bmatrix}\right\|_2^2.
\end{align*}
In short for any $\mathbf{u},\mathbf{f}\in\mathbb{C}^n$ we have
\begin{align}
\label{temp}
\begin{bmatrix}\mathbf{u}\\\bar{\mathbf{u}}\end{bmatrix}^H\nabla^2\mathcal{L}_{\epsilon}(\mathbf{f})\begin{bmatrix}\mathbf{u}\\\bar{\mathbf{u}}\end{bmatrix}\le L\left\|\begin{bmatrix}\mathbf{u}\\\bar{\mathbf{u}}\end{bmatrix}\right\|_2^2,
\end{align}
with $L=\left\|\mathbf{A}^H\mathbf{A}\right\|$. Combining the latter identity with a Wirtinger derivate version of Taylor's approximation theorem (e.g.~see \cite[Section 6]{WF}) we have that for any $\mathbf{f},\mathbf{f}^{+}\in\mathbb{C}^n$
\begin{align*}
\mathcal{L}_{\epsilon}(\mathbf{f}^{+})=&\mathcal{L}_{\epsilon}(\mathbf{f})+\begin{bmatrix}\nabla \mathcal{L}_{\epsilon}(\mathbf{f})\\\\\overline{\nabla \mathcal{L}_{\epsilon}(\mathbf{f})}\end{bmatrix}^H\begin{bmatrix}\mathbf{f}^{+}-\mathbf{f}\\\\\overline{\mathbf{f}^{+}-\mathbf{f}}\end{bmatrix}+\frac{1}{2}\begin{bmatrix}\mathbf{f}^{+}-\mathbf{f}\\\\\overline{\mathbf{f}^{+}-\mathbf{f}}\end{bmatrix}^H\left(\int_0^1\nabla^2\mathcal{L}_{\epsilon}\left(\mathbf{f}+t(\mathbf{f}^{+}-\mathbf{f})\right)\right)\begin{bmatrix}\mathbf{f}^{+}-\mathbf{f}\\\\\overline{\mathbf{f}^{+}-\mathbf{f}}\end{bmatrix},\\
\le&\mathcal{L}_{\epsilon}(\mathbf{f})+\begin{bmatrix}\nabla \mathcal{L}_{\epsilon}(\mathbf{f})\\\\\overline{\nabla \mathcal{L}_{\epsilon}(\mathbf{f})}\end{bmatrix}^H\begin{bmatrix}\mathbf{f}^{+}-\mathbf{f}\\\\\overline{\mathbf{f}^{+}-\mathbf{f}}\end{bmatrix}+\frac{L}{2}\left\|\begin{bmatrix}\mathbf{f}^{+}-\mathbf{f}\\\\\overline{\mathbf{f}^{+}-\mathbf{f}}\end{bmatrix}\right\|_2^2.
\end{align*}
Now plugging $\mathbf{f}^{+}=\hat{\mathbf{f}}_{\tau+1}=\hat{\mathbf{f}}_\tau-\mu\nabla \mathcal{L}_{\epsilon} (\hat{\mathbf{f}}_\tau)$ with $\mu=1/L$ and $\mathbf{f}=\hat{\mathbf{f}}_\tau$ in the above identity we conclude that
\begin{align*}
\mathcal{L}_{\epsilon}(\hat{\mathbf{f}}_{\tau+1})\le& \mathcal{L}_{\epsilon}(\hat{\mathbf{f}}_{\tau})-\mu\left(1-\frac{\mu L}{2}\right)\left\|\begin{bmatrix}\nabla\mathcal{L}_{\epsilon}(\hat{\mathbf{f}}_\tau)\\\\\overline{\nabla\mathcal{L}_{\epsilon}(\hat{\mathbf{f}}_\tau)}
\end{bmatrix}\right\|_2,\\
=&\mathcal{L}_{\epsilon}(\hat{\mathbf{f}}_{\tau})-\frac{1}{2L}\left\|\begin{bmatrix}\nabla\mathcal{L}_{\epsilon}(\hat{\mathbf{f}}_\tau)\\\\\overline{\nabla\mathcal{L}_{\epsilon}(\hat{\mathbf{f}}_\tau)}
\end{bmatrix}\right\|_2^2,\\
=&\mathcal{L}_{\epsilon}(\hat{\mathbf{f}}_{\tau})-\frac{1}{L}\left\|\nabla\mathcal{L}_{\epsilon}(\hat{\mathbf{f}}_\tau)\right\|_2^2.
\end{align*}
Rearranging the above inequality we conclude that
\begin{align*}
\left\|\nabla\mathcal{L}_{\epsilon}(\hat{\mathbf{f}}_\tau)\right\|_2^2\le L\left(\mathcal{L}_{\epsilon}(\hat{\mathbf{f}}_{\tau})-\mathcal{L}_{\epsilon}(\hat{\mathbf{f}}_{\tau+1})\right).
\end{align*}
Now summing over $\tau=1,2,\ldots, T$ we conclude that
\begin{align*}
\sum_{\tau=0}^T \left\|\nabla\mathcal{L}_{\epsilon}(\hat{\mathbf{f}}_\tau)\right\|_2^2\le L\left(\mathcal{L}_{\epsilon}(\hat{\mathbf{f}}_{0})-\mathcal{L}_{\epsilon}(\hat{\mathbf{f}}_{T+1})\right)\le L\left(\mathcal{L}_{\epsilon}(\hat{\mathbf{f}}_{0})-\mathcal{L}_{\epsilon}(\hat{\mathbf{f}})\right).
\end{align*}
Note that the above expression holds for any $\epsilon$. Thus taking the limit of both sides as $\epsilon\rightarrow 0$ we conclude that
\begin{align*}
\sum_{\tau=0}^T \left\|\nabla\mathcal{L}(\hat{\mathbf{f}}_\tau)\right\|_2^2\le L\left(\mathcal{L}(\hat{\mathbf{f}}_{0})-\mathcal{L}(\hat{\mathbf{f}})\right).
\end{align*}
Since the series above converges we must have $\underset{\tau\rightarrow\infty}{\lim} \left\|\nabla \mathcal{L}(\hat{\mathbf{f}}_\tau)\right\|_F\rightarrow 0$. Furthermore, we have
\begin{align*}
\min_{\tau\in\{0,1,2,\ldots,T\}} \left\|\nabla \mathcal{L}(\hat{\mathbf{f}}_\tau)\right\|_F^2\le\frac{1}{T+1}\sum_{\tau=0}^T \left\|\nabla\mathcal{L}(\hat{\mathbf{f}}_\tau)\right\|_2^2\le \frac{L\left(\mathcal{L}(\hat{\mathbf{f}}_{0})-\mathcal{L}(\hat{\mathbf{f}})\right)}{T+1},
\end{align*}
proving the identities \eqref{statement} and concluding the proof.

\end{document}